\documentclass[12pt,preprint]{aastex}
\usepackage{graphicx}
\begin{document}

\title{The Energetics of Launching the Most Powerful Jets in Quasars: A Study of 3C82}
\author{Brian Punsly\altaffilmark{1}, Gary J. Hill\altaffilmark{2,3}, Paola Marziani\altaffilmark{4}, Preeti Kharb \altaffilmark{5}, Marco Berton\altaffilmark{6,7}, Luca Crepaldi\altaffilmark{8}, Briana L. Indahl \altaffilmark{2}, Greg Zeimann\altaffilmark{9}}
 \altaffiltext{1}{1415 Granvia Altamira, Palos Verdes Estates CA, USA
90274: ICRANet, Piazza della Repubblica 10 Pescara 65100, Italy and
ICRA, Physics Department, University La Sapienza, Roma, Italy,
brian.punsly@cox.net}
\altaffiltext{2}{Department of Astronomy, University of Texas at Austin, Austin, TX 78712, USA }
\altaffiltext{3}{McDonald Observatory, University of Texas at Austin, Austin, TX 78712, USA }
\altaffiltext{4}{INAF, Osservatorio Astronomico
di Padova, Italia}
\altaffiltext{5}{National Centre for Radio Astrophysics,
Tata Institute of Fundamental Research, Post Bag 3, Ganeshkhind,
Pune 411007, India}
\altaffiltext{6}{Finnish Centre for Astronomy with ESO (FINCA), University of Turku, Vesilinnantie 5, FI-20014 University of Turku, Finland}
\altaffiltext{7}{Aalto University Mets{\"a}hovi Radio Observatory, Mets{\"a}hovintie 114, FI-02540 Kylm{\"a}l{\"a}, Finland}
\altaffiltext{8}{Dipartimento di Fisica e Astronomia``G. Galilei", Universit\`a di Padova, Vicolo dell'Osservatorio 3, 35122 Padova, Italy}
\altaffiltext{9}{Hobby Eberly Telescope, University of Texas, Austin, Austin, TX, 78712, USA}

\begin{abstract}
3C 82 at a redshift of 2.87 is the most distant 3C (Third Cambridge Catalogue) quasar. Thus, it is a strong candidate to have the most luminous radio lobes in the Universe. 3C 82 belongs to the class of compact steep spectrum radio sources. We use single dish and interferometric radio observations in order to model the plasma state of these powerful radio lobes. It is estimated that the long-term time-averaged jet power required to fill these lobes with leptonic plasma is $\overline{Q} \approx 2.66 \pm 1.33 \times 10^{47} \rm{ergs/sec}$, among the largest time averaged jet powers from a quasar. Positing protonic lobes is not tenable since they would require two orders of magnitude more mass transport to the lobes than was accreted to the central black hole during their formation. The first high signal to noise optical spectroscopic observation obtained of this object indicates that there is a powerful high ionization broad line wind with a kinetic power $\sim 10^{45} \rm{ergs/sec}$ and a velocity $\sim 0.01$c. We also estimate from the broad lines in 2018 and the UV continuum in three epochs spread out over three decades that the accretion flow bolometric luminosity is $L_{\rm{bol}} \approx 3.2-5.8 \times 10^{46} \rm{ergs/sec}$. The ratio of $\overline{Q}/L_{\rm{bol}}\approx 6.91 \pm 3.41$, is perhaps the largest of any known quasar. Extremely powerful jets tend to strongly suppress powerful winds of ionized baryonic matter. Consequently, 3C 82 provides a unique laboratory for studying the dynamical limits of the central engine of outflow initiation in quasars.
\end{abstract}

\section{Introduction}
Radio loud quasars are amongst the most powerful sustained events in the Universe. They arise from intense accretion of plasma onto supermassive black holes. They have three primary channels of emission. The most powerful tends to be the viscous dissipation of the infalling gas. This produces a large ultraviolet (UV) flux known as the characteristic``blue bump" that is the signature of a quasar \citep{mal83}. The next most energetic channel is the large scale jets of plasma that can extend hundreds of kpc to their termination in enormous radio lobes that are typically larger than the host galaxy. Thirdly, there are a variety of wind-like outflows that can reach outflow velocities of $\sim0.1$c \citep{wey91,wey97}. We have targeted 3C 82 for an exploratory investigation that can provide unique clues to the fundamental physics that interconnect these three channels of emitted power. 3C82 is an extremely powerful radio source, it has perhaps the most luminous radio lobes in the known Universe (within the uncertainty in the radio observations), a 151 MHz flux density of $\sim 10.5$ Jy at a redshift of z = 2.87. Surprisingly, in spite of its enormous jet power, this is a rarely studied object. To this end, we provide new radio imaging and the first high signal to noise ratio (SNR) optical spectrum.
\par 3C 82 with a small size projected on the sky plane, $\sim 11.3$ kpc is formally a subclass of radio source known as a compact steep spectrum radio source (CSS).
The CSS sources are a particular class of small extragalactic radio source. They are intrinsically small and this distinguishes them from blazars which appear small because they are observed along the jet axis and the projection of the source on the sky plane is consequently foreshortened \citep{bar89}. The evidence that they are intrinsically small is based on the lack of signatures of the Doppler variability or enhancement associated with the near pole-on view of a relativistic jet, no strong variability in the optical or radio and no elevated optical polarization \citep{bla79,lin85,ode98}. The other intrinsically small sources are gigahertz peaked sources (GPS) and high frequency peaked sources (HFP) \citep{ode98,ori08}. All three types of radio source appear to have synchrotron self-absorbed (SSA) powerlaw spectra, in which the spectral peak frequencies ($\nu_{\rm{peak}}$) for CSS, GPS and HFP sources are $\sim 100$ MHz, $\sim 1$ GHz and $>5$ GHz, respectively \citep{ode98,ori08}. The HFP sources with their high frequency turnover are actually a mix of intrinsically small sources and blazars \citep{ori08}. An empirical law was found to hold for many sources, $\nu_{\rm{peak}}\sim L_{\rm{sky}}^{-0.65}$, where the projected size on the sky plane is $L_{\rm{sky}}$ \citep{ode98}. The CSS sources are the largest of these small radio sources, but are still at a size less than the galactic dimension and this can be taken as a working definition \citep{ode98}. The CSS sources could be small due to two possible effects. They could be frustrated by the denser galactic environment, but in general it is believed that most are in the early stages of an evolutionary sequence in which the CSS sources are younger versions of the larger radio sources ($L_{\rm{sky}}> 50 \rm{kpc}$), the Fanaroff-Riley I and II (FRI and FRII) morphology radio sources discovered at low frequency and low resolution \citep{fr74,ode98}.
\par Quasars such as 3C 82 are believed to be the manifestation of the thermal emission generated by the viscous dissipation associated with the enormous shear forces of ionized gas that spirals inwards towards a central supermassive black hole \citep{mal83}. This central engine is shrouded in mystery since it is many orders of magnitude smaller than the resolution limits of modern telescopes. So astronomers must grapple with indirect clues to constrain the exotic physics of the quasar. The hot, shearing gas that characterizes the quasar central engine is in a regime that cannot be replicated in an Earth based laboratory. So we must resort to extrapolating better known phenomena, that are somewhat similar, in order to characterize the physics of the central engine. The best studied physical systems with comparable heating and ionization states are the solar atmosphere and solar wind. However, the plasma physics in these physical systems is far from being well understood and is an active field of astrophysical research. There is uncertainty in the physics associated with heating wind plasma, the time evolution of magnetic flux and flares as well as how the wind and coronal mass ejections are launched \citep{bau13,mal09,thr12,yam07}. At the most basic level, this is tantamount to an uncertainty about which microscopic terms and macroscopic collective phenomena to include in the algorithms and equations that govern the time evolution of the plasma. This uncertainty in the fundamental plasma equations is carried over with extrapolations to the quasar accretion flow onto the black hole. The situation is even more uncertain in quasars which have the most intense known shearing forces in the Universe. Adding large scale magnetic flux to the mix in order to launch the jets in radio quasars such as 3C 82 adds even more uncertainty and speculation to models \citep{pun15}. One of our indirect clues to the physical situation within the central engine is the maximum power that can be transported by the jet. 3C 82 is a unique laboratory for investigating the physics of jet launching from quasars, since it might be a limiting case for the maximum sustainable jet power. We take this study of 3C82 as an opportunity to investigate the maximum long-term time-averaged jet power of a quasar, $\overline{Q}$, and the maximum magnitude of jet power relative to the bolometric thermal luminosity of a quasar accretion flow, $L_{\rm{bol}}$, $R=\overline{Q}/L_{\rm{bol}}$.

\par In constraining $\overline{Q}$ for 3C82, a major issue is the small size projected on the sky plane, $\sim 11.3$ kpc. The standard estimates of $\overline{Q}$ may not apply to CSS sources or jets propagating in dense environments, in general \citep{bar96,wil99}. These estimates assume relaxed lobes from a classical double radio source, i.e. with lobes outside of the host galaxy. If the lobes are constrained by an ambient galactic medium (which is of higher density than the medium which surrounds lobes on super-galactic dimensions) then they will be ``over-luminous" and these methods will over-estimate $\overline{Q}$. Thus, a different approach is required for an accurate $\overline{Q}$ estimate in CSS quasars. The method that we implement to estimate $\overline{Q}$ has been successful in estimating power in discrete ejections in gamma ray bursts, galactic black holes and other quasars \citep{rey09,rey20,pun12,pun19}. It requires that the spectrum appears as a power-law with a SSA turnover. In this circumstance, one can obtain an estimate of the spatial dimension of the region that creates the preponderance of the lobe radio luminosity, a dimension that can be much less than the unresolved image size inferred from interferometric observations \citep{rey09}. This derived dimension, the luminosity and the spectral index can be used to estimate the energetics of the ejection if the bulk flow Doppler factor of the plasmoid, $\delta$, is constrained from observations \citep{pun12,lig75}.
\par The paper is organized as follows. Section 2 describes the new optical observations and provides a detailed discussion of the luminosity and asymmetric shapes of the broad emission lines (BELs). From that we estimate $L_{\rm{bol}}$ from our new 2018 spectrum. In section 3, we describe new radio images produced from data in the archives of the Very Large Array (VLA) and the Jansky Very Large Array (JVLA). In sections 4-7, we derive energy estimates for the radio lobes and convert this to $\overline{Q}$. In Section 8, we explore the energy flux of the high ionization wind based on the UV spectrum.
Throughout this paper, we adopt the
following cosmological parameters: $H_{0}$=69.6 km s$^{-1}$Mpc$^{-1}$, $\Omega_{\Lambda}=0.714$ and $\Omega_{m}=0.286$ and use Ned Wright’s Javascript
Cosmology Calculator \footnote{http://www.astro.ucla.edu/~wright/CosmoCalc.html} \citep{wri06}.
\begin{figure}
\begin{center}
\includegraphics[width= 0.85\textwidth]{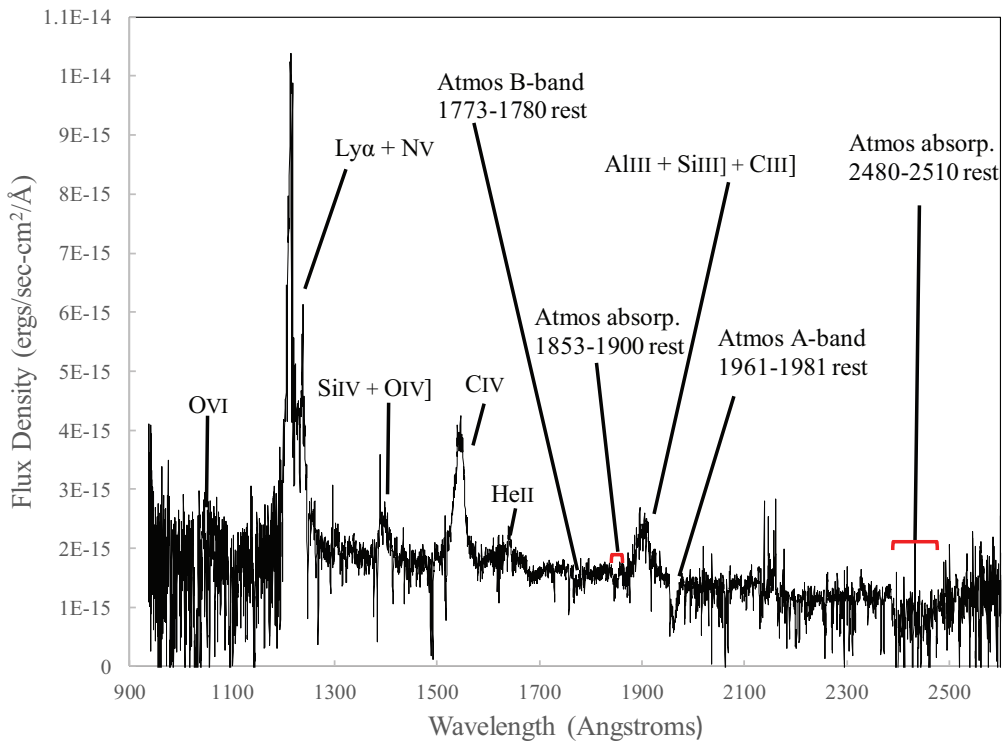}

\caption{The HET LRS2 spectrum of 3C 82 showing the prominent emission lines. The data are presented in the cosmological frame of reference of the quasar (for both wavelength and flux density) in order to get a clear picture of the atomic origin of the emission lines and the magnitude of the quasar luminosity. The spectrum was calibrated as described in the text. It is corrected for Galactic extinction, but not for telluric absorption. Prominent telluric features are marked. For the analysis, the part of the spectrum around the $\lambda~1900$ complex was corrected for telluric absorption as discussed in the text.}
\end{center}
\end{figure}

\section{The Ultraviolet Spectrum}
There exist two previous spectra of 3C 82, but the SNR are too low to establish the nature of the BELs \citep{raw89,sem04}. 3C 82 being at such a high redshift, z=2.87, is extremely faint, $m_{V}=21.0$. Thus, a large telescope is needed to produce a high quality spectrum. We used the new Low Resolution Spectrograph 2 (LRS2, Chonis et al. 2016, Hill et al. 2020 in prep.) on the upgraded 10m Hobby-Eberly Telescope (HET, Hill et al. 2018) to obtain a spectrum on November 7, 2018 UT. We used both units of the integral field spectrograph, LRS2-B and LRS2-R. Each unit is fed by an integral field unit with 6x12 square arcsec field of view, 0.6 arcsec spatial elements. and full fill-factor, and has two spectral channels. LRS2-B has two channels, UV (3700\AA ~- 4700\AA) and Orange (4600\AA ~- 7000\AA), observed simultaneously. Similarly, LRS2-R has two channels, Red (6500\AA ~- 8470\AA) and Far-red (8230\AA ~- 10500\AA). The observation was split into two 30 minute exposures one for LRS2-B and one for LRS2-R. The image size was 1.5 arcsec. full-width half-maximum. The spectra from each of the four channels were reduced independently using the HET LRS2 pipeline, Panacea (Zeimann et al. 2020, in prep.).
The primary steps in the reduction process are bias subtraction, dark subtraction, fiber tracing, fiber wavelength evaluation, fiber extraction, fiber-to-fiber normalization,  source extraction, and flux calibration.  For more details on Panacea v0.1, which was used for the LRS2 data reductions within this paper, see the code documentation\footnote{Panacea v0.1 documentation can be found at https://github.com/grzeimann/Panacea/blob/master/README \_ v0.1.md} and Zeimann et al. (2020), in prep.
Normalization of the reduced spectra from the 4 channels is accomplished in two steps. The spectra from each unit are obtained simultaneously and are well normalized. Some adjustment of the UV and Orange channels is needed to account for the shift in the location of the object on the IFU due to differential atmospheric refraction. This can create an aperture correction for the UV channel due to wings of the point spread function falling off the IFU. The setup for 3C82 was accurate and the correction was negligible. The same was the case for the Red and Far-red channels. Since the HET illumination varies during the track, the separate observations with LRS2-B and -R need to be normalized, along with any changes of transparency during the track. When these effects were taken into account a small 5\% additional multiplicative offset was needed to normalize the spectra in the region of overlap (6500 to 7000\AA). Finally, the combined spectrum was placed on an absolute flux scale through
comparison with the g-band magnitude obtained from the acquisition camera (ACAM). Using Sloan Digital Sky Survey (SDSS) catalog stars in the ACAM field, the g-band transparency was measured to be 85\% and the magnitude of 3C82 was estimated to be g = 20.23. The LRS2 spectrum was normalized to this magnitude by convolving with the g-band filter profile and normalizing to the magnitude measured from the ACAM. After this process, we expect an absolute flux calibration accuracy of $\sim 10\%$.
\par The rest frame spectrum presented in Figure 1 has effective integration time of 30 minutes and was corrected for Galactic extinction. The best fit to the extinction values in the NASA Extragalactic Database (NED) in terms of \citet{car89} models is $A_{V}=0.51$ and $R_{V}=2.8$. The first thing to note is that the UV flux density at 1350~\AA\, is approximately 20\% higher than in the noisy November 2002 and January 1988 spectra \citep{sem04,raw89}. These differences may be accounted for in calibration uncertainty, such as slit losses in the earlier data or some modest variability. But, there is no sign of the extreme optical variability indicative of a (nearly polar) blazar line of sight.
\subsection{The Broad Emission Lines}
\begin{table}
\caption{Ultraviolet Broad Emission Line Fits}
{\tiny\begin{tabular}{cccccccccccc} \tableline\rule{0mm}{3mm}
 Vacuum &  Line & Red VBC & Red VBC & Red VBC & BLUE & BLUE & BLUE & BC & BC & Total BEL \\
Wavelength &   &  Peak\tablenotemark{c} & FWHM & Luminosity & Peak\tablenotemark{c} & FWHM &Luminosity & FWHM &Luminosity & Luminosity  \\
(Angstroms) &    &  km~s$^{-1}$ &  km~s$^{-1}$ & ergs~s$^{-1}$ & km~s$^{-1}$ &km~s$^{-1}$ & ergs~s$^{-1}$ & km~s$^{-1}$  & ergs~s$^{-1}$ & ergs~s$^{-1}$  \\
\tableline \rule{0mm}{3mm}
1215.67 & Ly $\alpha$ & $3713$ & $8047\pm 491$ &  $5.20 \times 10^{44}$& $-1646$ &$4993\pm430$  & $4.41 \times 10^{44}$ & $2368 \pm 1071$& $4.25 \times 10^{44}$ & $1.39 \times 10^{45}$ \\
1240.8 & NV & $3713$ & $8047\pm 491$ &  $1.41 \times 10^{44}$& $-1644$  & $5821\pm 465 $ &$7.18 \times 10^{43}$ & $2368 \pm 1071$& $2.95 \times 10^{43}$ & $2.42 \times 10^{44}$ \\
$\sim 1400$ & SiIV+OIV] & \tablenotemark{a}& \tablenotemark{a} &  \tablenotemark{a}& \tablenotemark{a} & \tablenotemark{a}  & \tablenotemark{a} & \tablenotemark{a}& \tablenotemark{a}& $1.07\times 10^{44}$ \\
1549.06 & CIV & $1332$ & $7163\pm 530$ &  $7.61 \times 10^{43}$& $-2618$  & $6070\pm365$ & $1.78 \times 10^{44}$ & $3448 \pm 166$ & $2.47 \times 10^{44}$ & $5.01 \times 10^{44}$ \\
1640.36 & HeII & $3116$ & $7163\pm 530$ &  $7.17 \times 10^{43}$& $-5250 $  & $6070\pm365$ & $6.15 \times 10^{43}$ & $4702 \pm 689$ & $5.25\times 10^{43}$ & $1.86\times 10^{44}$ \\
1854.47 & AlII & \tablenotemark{a} & \tablenotemark{a} & \tablenotemark{a} & \tablenotemark{a} & \tablenotemark{a} & \tablenotemark{a}  & \tablenotemark{a} & $9.08 \times 10^{42}$ & $9.08 \times 10^{42}$ \\
1862.79 & AlIII & \tablenotemark{a}& \tablenotemark{a} & \tablenotemark{a} & \tablenotemark{a} & \tablenotemark{a} & \tablenotemark{a}  & \tablenotemark{a}  & $7.28 \times 10^{42}$ & $7.28 \times 10^{42}$ \\
1892.03 & SiIII] & $845$ & $5335\pm 2310$ & $4.20 \times 10^{43}$ & \tablenotemark{b} & \tablenotemark{b} & \tablenotemark{b}  & $2808\pm 594$  & $4.94 \times 10^{43}$ & $9.13 \times 10^{43}$ \\
1908.73 & CIII] & $891$ & $5335\pm 2310$ & $5.29 \times 10^{43}$ & \tablenotemark{b} & \tablenotemark{b} & \tablenotemark{b}  & $2296 \pm 582$  & $7.41 \times 10^{43}$ & $1.27 \times 10^{44}$ \\ \hline
\end{tabular}}
\tablenotetext{a}{Insufficient SNR for an accurate decomposition.}
\tablenotetext{b}{Not detected.}
\tablenotetext{c}{Peak of the fitted Gaussian component in km/sec relative to the quasar rest frame. A positive value is a redshift.}
\end{table}
The BELs in Figure 1 are of particular interest because they provide clues to the nature of the gas near the quasar and the luminosity of the quasar. We provide a standard three Gaussian component decomposition of the BELs, the broad component, BC, \citep[also called the intermediate broad line or IBL;][]{bro94}, the redshifted  VBC \citep[very broad component following][]{sul00} and a blueshifted excess \citep{bro96,mar96,sul00}. In summary, there are three broad components defined by their velocity relative to the quasar rest frame. The redshifted broad component is the broadest of the three components. It will be abbreviated in the following with a prefix ``red" as ``the red VBC." The blue shifted broad excess is designated as BLUE. The component that is close to the quasar rest frame, the IBL/BC, will have no prefix to the BC and just be called the ``BC" in the following. These designations are used in Table 1 to describe the three component decompositions of the broad lines.

The decompositions are shown in Figure 2, after continuum subtraction. The BC are the black Gaussian profiles in Figure 2. The red VBC is shown in red and BLUE is shown in blue. Only the sum of the three components is shown for both NV$\lambda$1240 (contaminating the red wing of Ly$\alpha$) and HeII$\lambda$1640 (merging with red wing of CIV and creating the appearance of a flat-topped profile). In the $\chi^2_\nu$ minimization fit, the decomposition was carried out in a consistent way, with similar initial guesses of line shift and width values for the three components in CIV, HeII, Ly$\alpha$ and NV, derived from the decomposition of the CIV profile. Their relative intensity was allowed to vary freely (i.e., the relative intensity of the three components is not constrained by the CIV decomposition.). With this approach it was possible to obtain a minimum $\chi^{2}_\nu$ fit that leaves no significant residuals in the decomposition of the main blends (see Figure 2) for both CIV+HeII and Ly$\alpha$ + NV fit. The Ly$\alpha$+NV blend is especially problematic, as the blue side of the line is contaminated by the Ly$\alpha$\ forest. The absorptions may have eaten away most of the flux of the BLUE, whose intensity is therefore especially uncertain. In quasars with broader emission lines (such as the Population B sources described below), the CIV+HeII blend flat topped appearance (e.g., Fine et al. 2010) can be explained by the blending of the CIV redwing and HeII blue wing without the assumption of any additional emission \citep{mar10}. The flat top implies that the HeII BC is weak with respect to the HeII BLUE and red VBC: it is not possible to use a scaled CIV profile to model HeII. A similar condition is apparently occurring also for NV.
\par The CIII] complex is affected by telluric absorption. We used the standard star, HD 84937, which was observed on the same night, to correct for this. We could not fully correct for the A-band absorption resulting in the dip on the red side of the complex (indicated in the bottom right hand panel of Figure 2). Regions affected by absorption lines are avoided in fitting. The fit in Figure 2 was obtained by restricting the fitting range to the intervals  1795--1845, 1849–1950, 1965–1980, 1990–2010 \AA.  Table 1 labels AlIII with insufficient SNR for an accurate decomposition. We report the total luminosity but note that the AlIII doublet is potentially affected by telluric residuals. The SNR is adequate to expose the complexity of the $\sim 1900$~\AA\, blend, and to reveal CIII] and SiIII] profiles typical of quasars with very broad low ionization emission lines (see the discussion of Population B sources, below). There is a prominent BC, red VBC and no BLUE detected in these lines.
\par These line decompositions (shown in Figure 2) are described quantitatively in Table 1. The table is organized as follows. The line designation is defined in the first two columns. The next three columns define the properties of the Gaussian fit to the red VBC, the shift of the Gaussian peak relative to the vacuum wavelength in km/sec, followed by the FWHM and line luminosity. Columns (6)-(8) are the same for the BLUE. The BC FWHM and luminosity are columns (9) and (10). The last column is the total luminosity of the BEL.
\par The broad line decomposition has a physical context. In order to explore this, we note that there are two different, useful, ways of segregating the quasar population. One, is to split the population into radio quiet and radio loud quasars and the other is to split the quasars into Population A or B \citep{sul00}: Population A (H$\beta$ FWHM $<4000$ km/sec) and Population B (defined by H$\beta$ FWHM $>4000$ km/sec). The CIV BLUE has been found to be dominant relative to the CIV red VBC in radio quiet quasars. It tends to be less prominent in radio loud quasars and can be completely absent \citep{ric02,pun10}. The trend was shown to be deeper than just the radio loud - radio quiet dichotomy, but related more to the Eddington accretion rate onto the central supermassive black hole. The Population B quasars include most of the radio loud quasars and typically have very large black hole masses and low Eddington ratios: $R_{\rm{Edd}} \equiv L_{\rm{bol}}/L_{\rm{Edd}}\sim 0.01-0.1$, where $L_{\rm{bol}}$ is the thermal bolometric luminosity of the accretion flow and $L_{\rm{Edd}}= 1.26 (M_{bh}/M_{\odot})\times 10^{38}~\rm{ergs~s^{-1}}$ is the Eddington luminosity expressed in terms of the central supermassive black hole mass, $M_{bh}$ \citep{sul07}. The Population A quasars are generally radio quiet and have typically higher Eddington ratios than Population B quasars \citep{sul07}. Population B CIV profiles tend to have less blue excess at their bases than Population A quasars, but this distinction is less pronounced for high luminosity Population B quasars \citep{sul17,ric02}. These patterns are a strong indication that the BLUE is related to a wind driven by radiative luminosity.
\par The physical insight provided by the Population A/B discussion, above, sheds light on the nature of the central engine of 3C82. In radio loud quasars, even though BLUE of CIV is often detectable, it is usually significantly weaker than the red VBC \citep{pun10}. Yet, the BLUE of the CIV broad line in 3C82 is 2.3 times as luminous as the red VBC. This is very extreme for a radio loud quasar. The implication is that 3C 82 has a high $L_{\rm{bol}}$ for a radio loud quasar and this is related to the strong BLUE. In Section 8, we interpret this in terms of an out-flowing high ionization wind that is typically found in high Eddington rate quasars.

\begin{figure*}
\begin{center}
\includegraphics[width= 0.55\textwidth,angle =0]{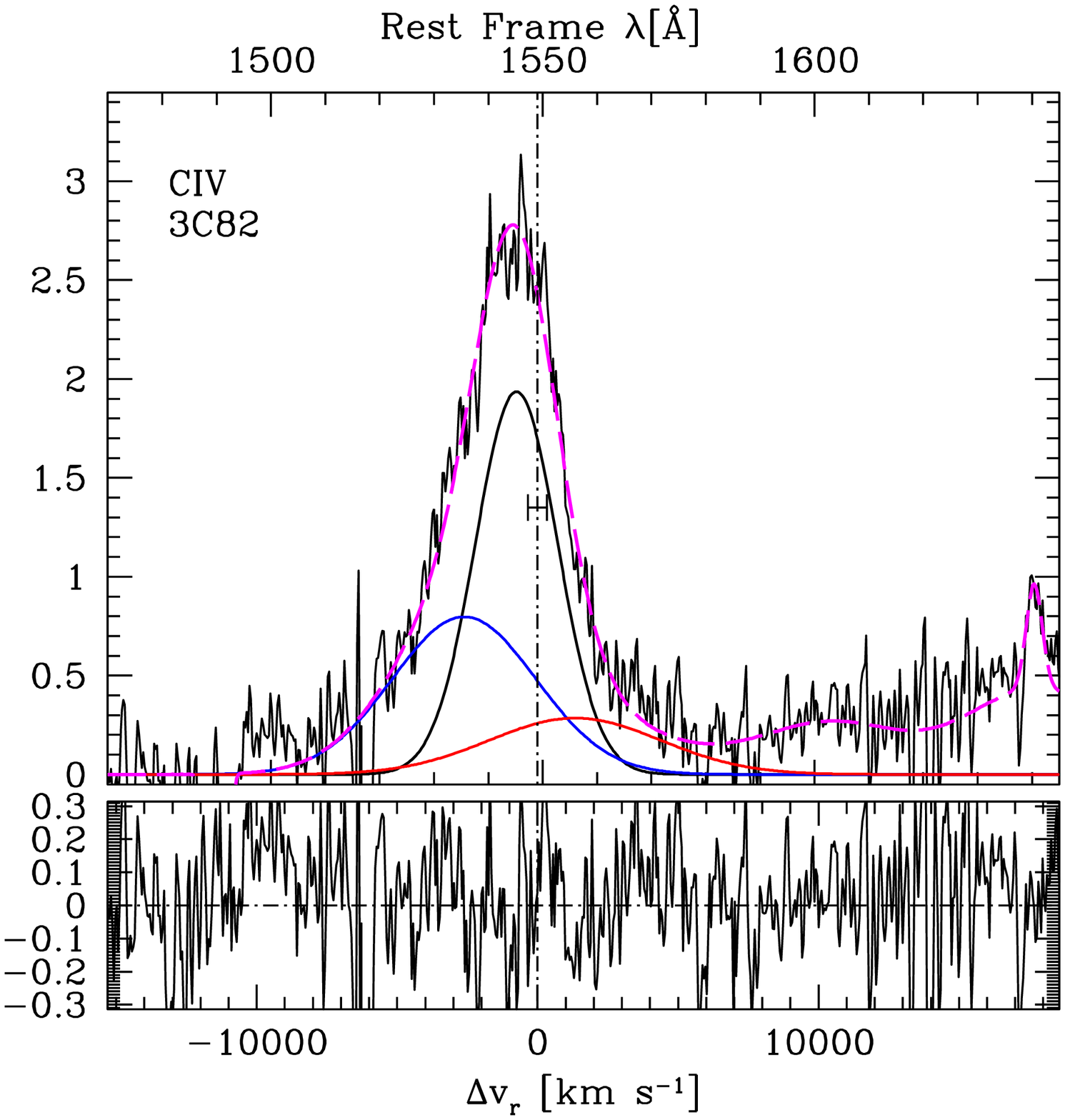}
\includegraphics[width= 0.45\textwidth,angle =0]{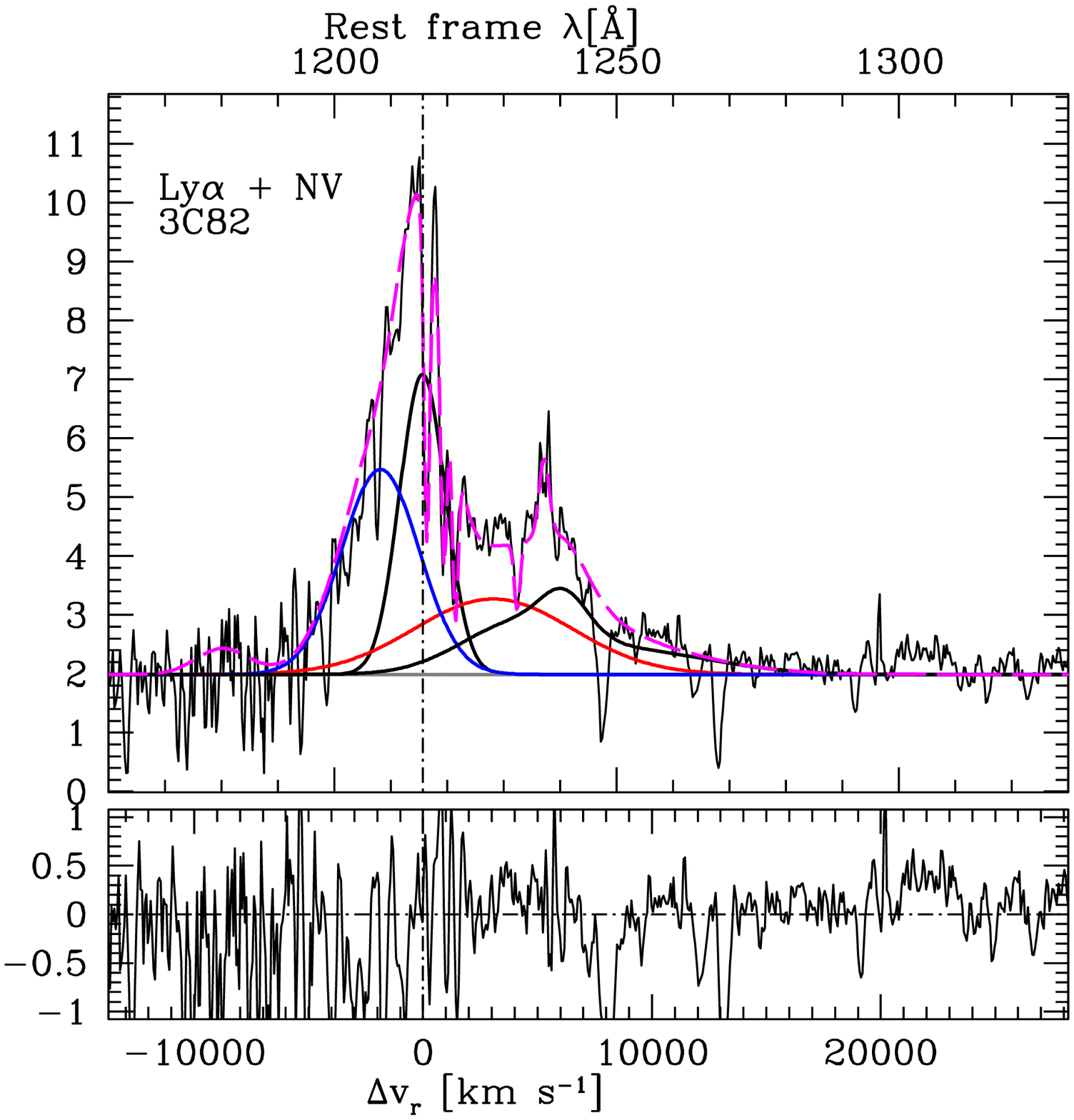}
\includegraphics[width= 0.45\textwidth,angle =0]{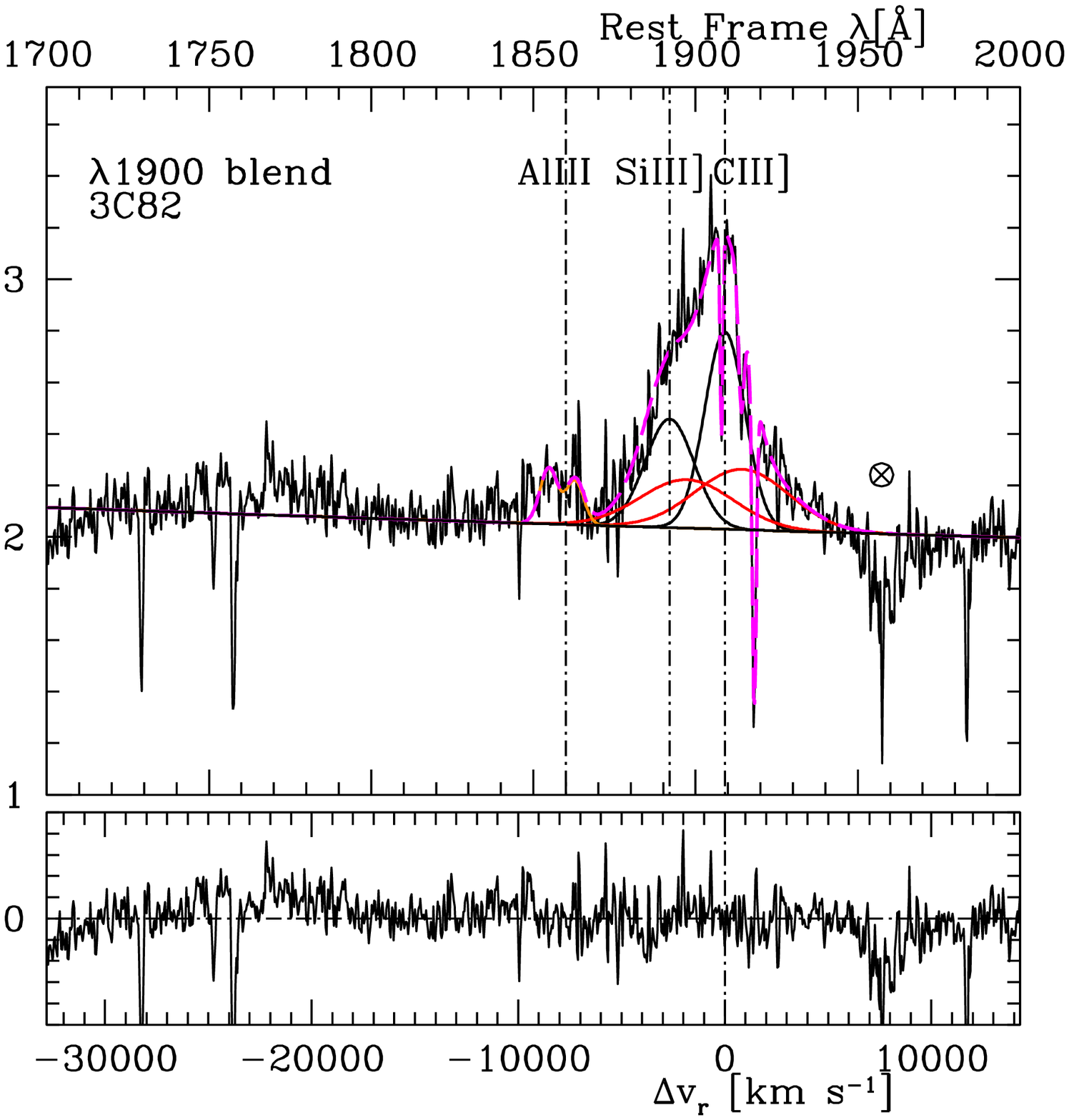}

\caption{\footnotesize{The three prominent broad line regions in Figure 1 fitted by the procedure discussed in the text. The results of the fitting process are presented in Table 1. The vertical axes are the flux density in units of $10^{-15}$ ergs/sec/$\rm{cm}^{2}$/\AA\, in the quasar rest frame. The BLUE (blue), red VBC (red) and BC (black) are shown as separate components, while only the sum of the three components (in black) is shown for NV$\lambda$1240 (contaminating the red wing of Ly$\alpha$) and HeII$\lambda$1640 (merging with the red wing of CIV and creating the appearance of a flat-topped profile.) Note the existence of the red VBC in CIV and the much stronger BLUE in CIV. Also, note the strong SiIII] on the blue side of CIII] (see Table 1). The prominence of the BLUE is less in the lower ionization lines providing evidence of a high ionization outflow.}}
\end{center}
\end{figure*}

\subsection{The Bolometric Luminosity of the Accretion Flow} We wish to estimate $L_{\rm{bol}}$ in a manner that does not include
reprocessed radiation in the infrared from molecular clouds that are far from the active nucleus.
This would be double counting the thermal accretion emission that is
reprocessed at mid-latitudes \citep{dav11}. The most direct method is to use the UV continuum as a surrogate for $L_{\rm{bol}}$.
From the spectrum in Figure 1 and the formula expressed in terms of quasar cosmological rest frame wavelength,
$\lambda_{e}$ and spectral luminosity, $L_{\lambda_{e}}$, from \citet{pun16},
\begin{equation}
L_{\mathrm{bol}} \approx (4.0 \pm 0.7)\lambda_{e}L_{\lambda_{e}}(\lambda_{e} = 1350 \AA)\approx 4.9 \pm
0.9 \times 10^{46} \rm{ergs/s} \;.
\end{equation}
The bolometric correction was estimated from a comparison to HST composite spectra of quasars with $L_{\mathrm{bol}} \approx \times 10^{46} \rm{ergs/s}$ \citep{zhe97,tel02,lao97}. The estimate is 20\% lower if one uses the lower SNR spectra from 1988 and 2002.
One can also use the luminosity of the CIV BEL in Table 1 as a more indirect means of estimation,
\citep{pun16},
\begin{eqnarray}
&&L_{\mathrm{bol}}=(107 \pm 22) L(CIV)\approx 5.4\pm 1.1 \times 10^{46} \rm{ergs/s}:\,
\end{eqnarray}
The uncertainty in Equations (1) and (2) arises from the uncertainty in the Hubble
Space Telescope composite continuum level and the uncertainty in $L(CIV)$ and $L_{\lambda_{e}}(\lambda_{e} = 1350 \AA)$, respectively, added in quadrature \citep{tel02,pun18}.
\section{Radio Observations}
We reduce three archival radio observations, all of which are unpublished. The August 4, 1991 A-array, 6 minute, 8.4 GHz, X-band. observation of the VLA in the top left hand panel of Figure 3 is the most useful since it is resolved into a double radio source, (project code AE0081). There is also clear evidence of a jet entering the western lobe. The western lobe plus jet (eastern lobe) has a flux density of 94 (39) mJy. The jet-like feature has $\sim 1$ mJy of flux density and is undetected in our other observations. The western lobe appears to be elongated and larger.
\par The upper right hand panel of Figure 3 shows a 327 MHz VLBA (Very Long Baseline Array) image of the western lobe that was previously published and kindly provided by N. Kanekar \citep{kan13}. The high resolution image is shown again, here, in order to help us understand the brighter western lobe. There seems to be a modestly bright hot spot, the peak surface brightness in the western portion of the lobe. Comparing this to the X-band image in the top left hand frame seems to indicate a jet entering the lobe on the eastern side that bends abruptly to the west at the working surface of the lobe against the ambient environment, before terminating at the hot spot. This morphology is very common in high resolution images of Fanaroff-Riley II (FRII) radio lobes \citep{kha08,fer14}. We note that this is a very high resolution image with sparse u-v coverage and a significant fraction of the flux is not detected. Thus, the flux density of the components extracted from the 327 MHz VLBA image are effectively lower bounds on the actual 327 MHz flux density of the components.

\par The lower left hand panel of Figure is a JVLA observation from July 24 2015 in A-array it was 15 minutes in duration and provides our most sensitive imaging of any possible diffuse lobe emission (Project Code 15A-155). The observation was at L-band, 1.0075 GHz -2.0315 GHz. There are sixteen 64 MHz spectral windows (spw), they are labeled spw0 to spw15. We reduced and analyzed the data using the Common Astronomy Software Applications (CASA) version 5.0.0-218 and the standard JVLA data reduction pipeline version 5.0.0. In our maps we used a pixel size of 0.15 arcsec to properly sample the primary beam. We carried out 5 cycles of CLEAN algorithm and self calibration. The central spectral windows had a higher noise level likely due to the presence of some radio frequency interference, therefore we decided to discard them. We produced instead two maps close to the edges of the band, centered at 1.1035 and 1.9355 GHz, each one with a bandwidth of 64 MHz. The resolution was insufficient to resolve the two components in spw1. The overall size is $\approx 1.42$ arcsec which is $\approx 11.3$ kpc in our adopted cosmology \citep{kan13}. However, the clean beam size (major axis FWHM: 1.39 arcsec and minor axis FWHM: 1.06 arcsec) was sufficient to partially resolve the source in spw14 and that image is the one chosen to be presented in Figure 3. The total flux density of the spw1 (the second lowest frequency window) image is 1794 mJy and 908 mJy for spw14 (the second highest frequency window). The uncertainty in the flux density measurements is 5\% based on the VLA manual\footnote{located at https://science.nrao.edu/facilities/vla/docs/manuals/oss/performance/fdscale}, see also \citep{per13}. We proceeded to fit two Gaussian components to the spw14 image using the \texttt{imfit} task of CASA. The fitted western Gaussian component is brighter with 568 mJy and the eastern Gaussian component is 342 mJy. We attribute a larger uncertainty to the components, individually, than the total flux density, 10\%. In this analysis, all parameters, the peak intensity values, peak position values and component sizes were free to vary in our fitting process.

\begin{figure}
\includegraphics[width= 1\textwidth,angle =0]{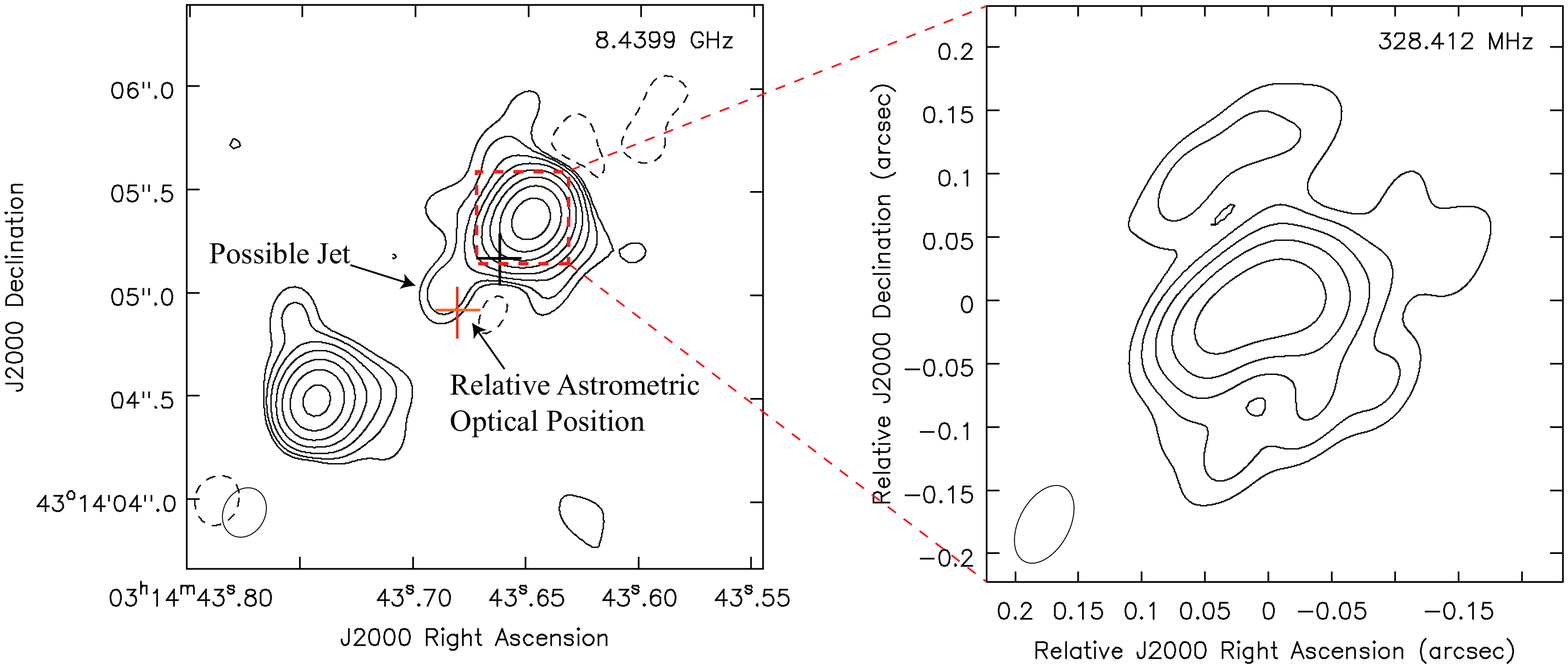}
\includegraphics[width= 0.5\textwidth,angle =0]{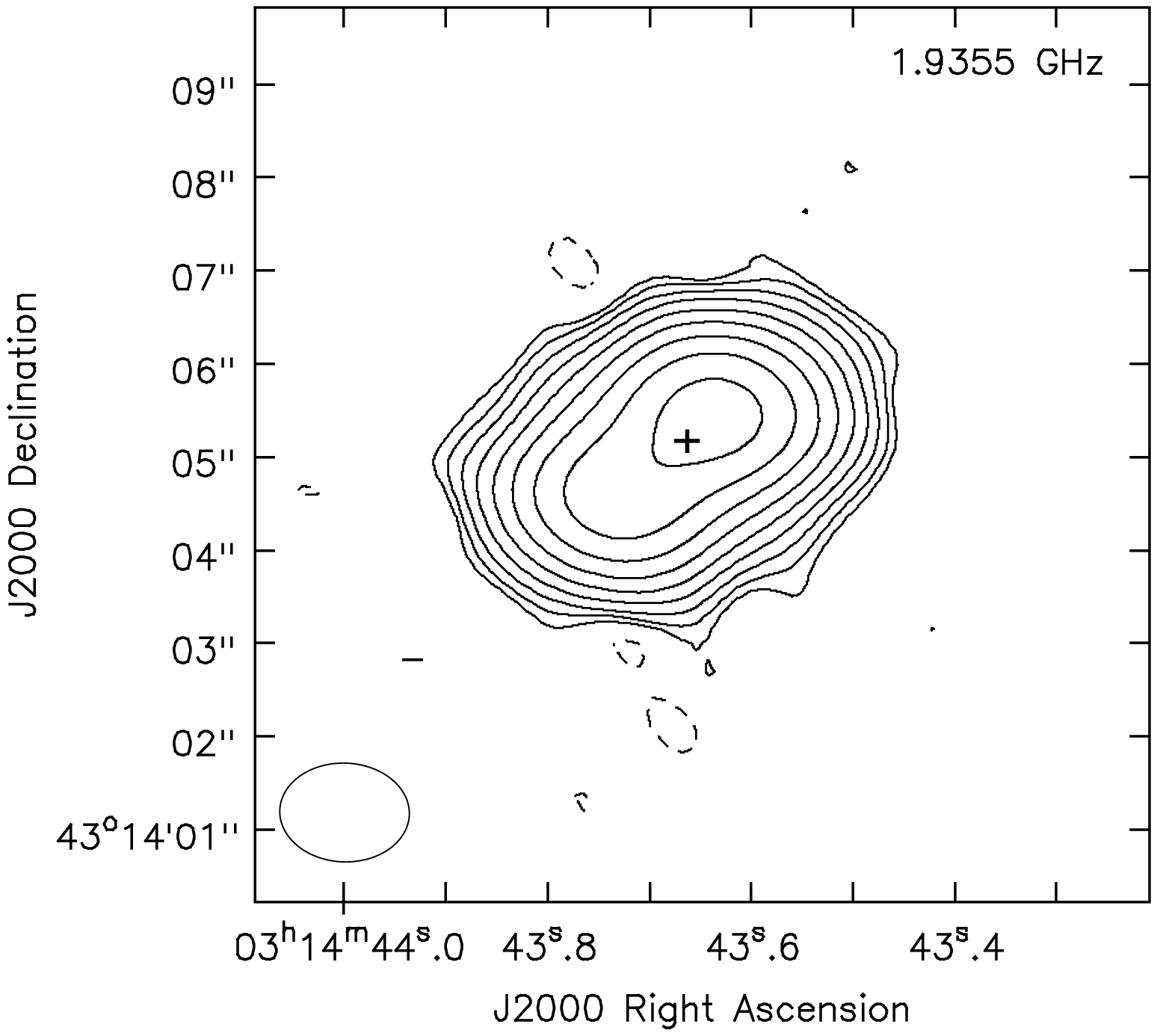}
\includegraphics[width= 0.5\textwidth,angle =0]{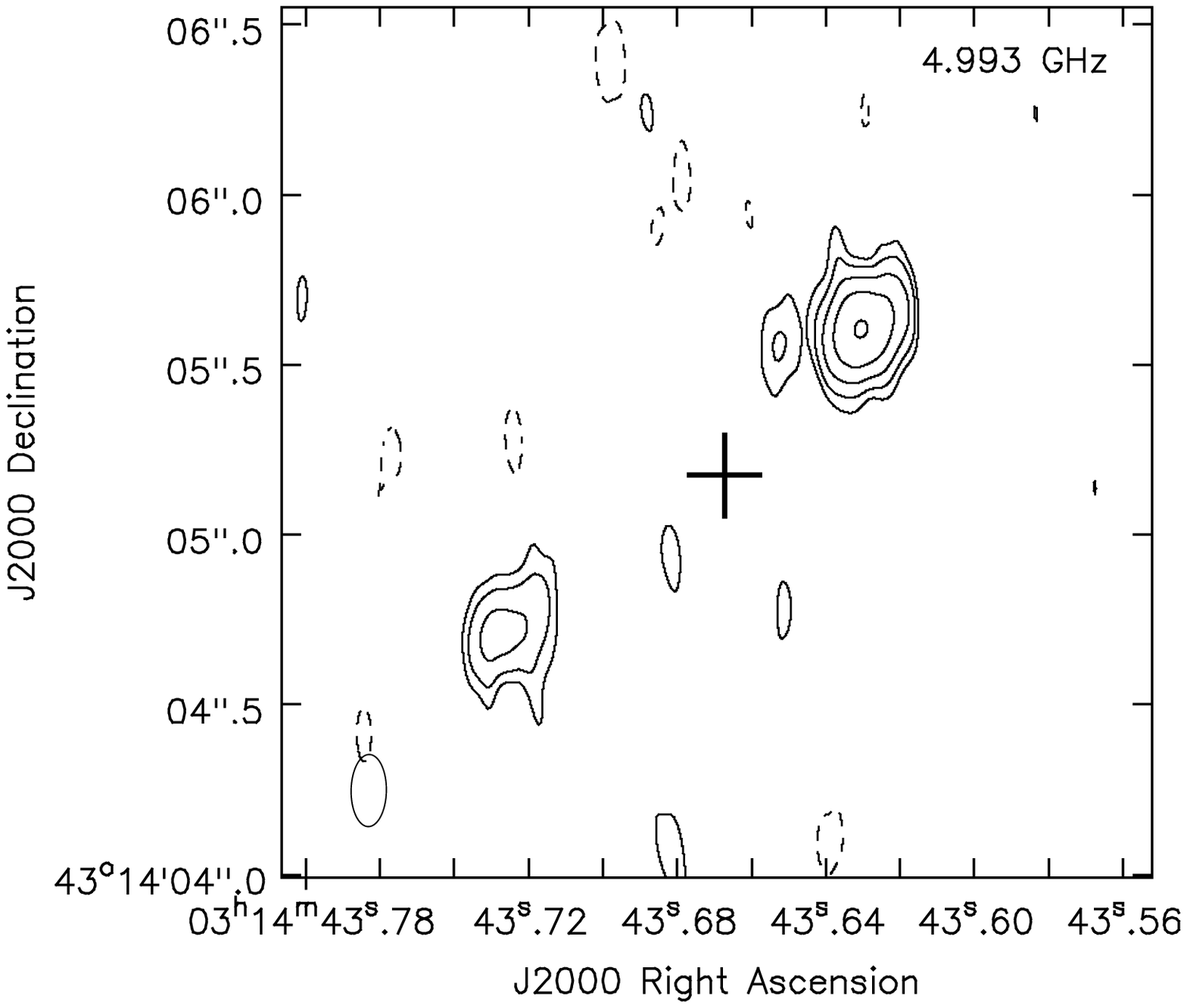}

\caption{\footnotesize{The top left (right) hand panel is the X-band 8.4 GHz (327 MHz) image from the VLA (VLBA). The VLBA image is a closeup of the region indicated in the red square. Due to poor positional accuracy resulting from the self-calibration procedure, the center of the red square is just an estimation. The 1.9355 GHz JVLA (5 GHz MERLIN) image is in the lower left (right) hand panel. The contour levels are at rms$\times$(-3, 3$\times$2$^n$), n $\in$ [0,x]. The rms, x values are: 0.13 mJy for VLA 8 GHz, x = 7; 10 mJy for VLBA 0.3 GHz: x = 4; 0.5 mJy for VLA 1.9 GHz, x = 8; 2.0 mJy for MERLIN 5 GHz, x = 4. The optical position is indicated by a cross of 1$\sigma$ error bars. However, the astrometry of the X band image is quite inaccurate and two crosses are displayed. The black cross is the position assuming the X-band astrometry is correct. The red cross is corrected for the reliable astrometry from the Merlin image and represents a more accurate optical position relative to the radio source (see the text for details).}}

\end{figure}

\par The lower right panel of Fig. 3 shows a much lower quality Multi-Element Radio-Linked Interferometer Network (MERLIN) image at 5 GHz for completeness. The data were taken on November 30, 1991 when the array had limited capabilities and only observed in single polarization (LL). The image required heavy u-v tapering in order to capture the bulk of the lobe flux. The western lobe plus jet (eastern lobe) has a flux density of 181 (63) mJy. Comparison of this image to the JVLA and VLA data above indicates that diffuse flux ($\sim 10-15$ mJy) is not detected (resolved out) in the eastern lobe in this MERLIN image. The main reason for showing the 5 GHz image is that it does not show any evidence of a radio core. The fact that the putative jet, tentatively detected at 8.4 GHz, is not seen in the MERLIN image is not unexpected. At 8.4 GHz this requires a dynamic range of 70 (second lowest contour level) to detect\footnote{The dynamic range is defined as peak flux density divided by the lowest positive contour above the image noise, where the noise is set by the most  negative contour level}. Since the lobe is likely steeper spectral index ($\alpha \approx 1.1$) than the jet, this indicates that a dynamic range of $>70$ is required at 5 GHz. Yet, the dynamic range of the MERLIN image is only 18 with similar restoring beam dimensions to the VLA, X-band image. We note that if the MERLIN data is imaged with full resolution ($ \sim$ 0.05" x 0.07" restoring beam) a faint, short, 0.1", elongation of the western lobe is seen at approximately the same PA as the putative jet at 8.4 GHz. This low quality MERLIN image is not used in this paper because it resolves out most of the diffuse emission of the lobes, the quantity of primary scientific interest for this study.

\begin{figure*}
\begin{center}
\includegraphics[width= 0.9\textwidth,angle =0]{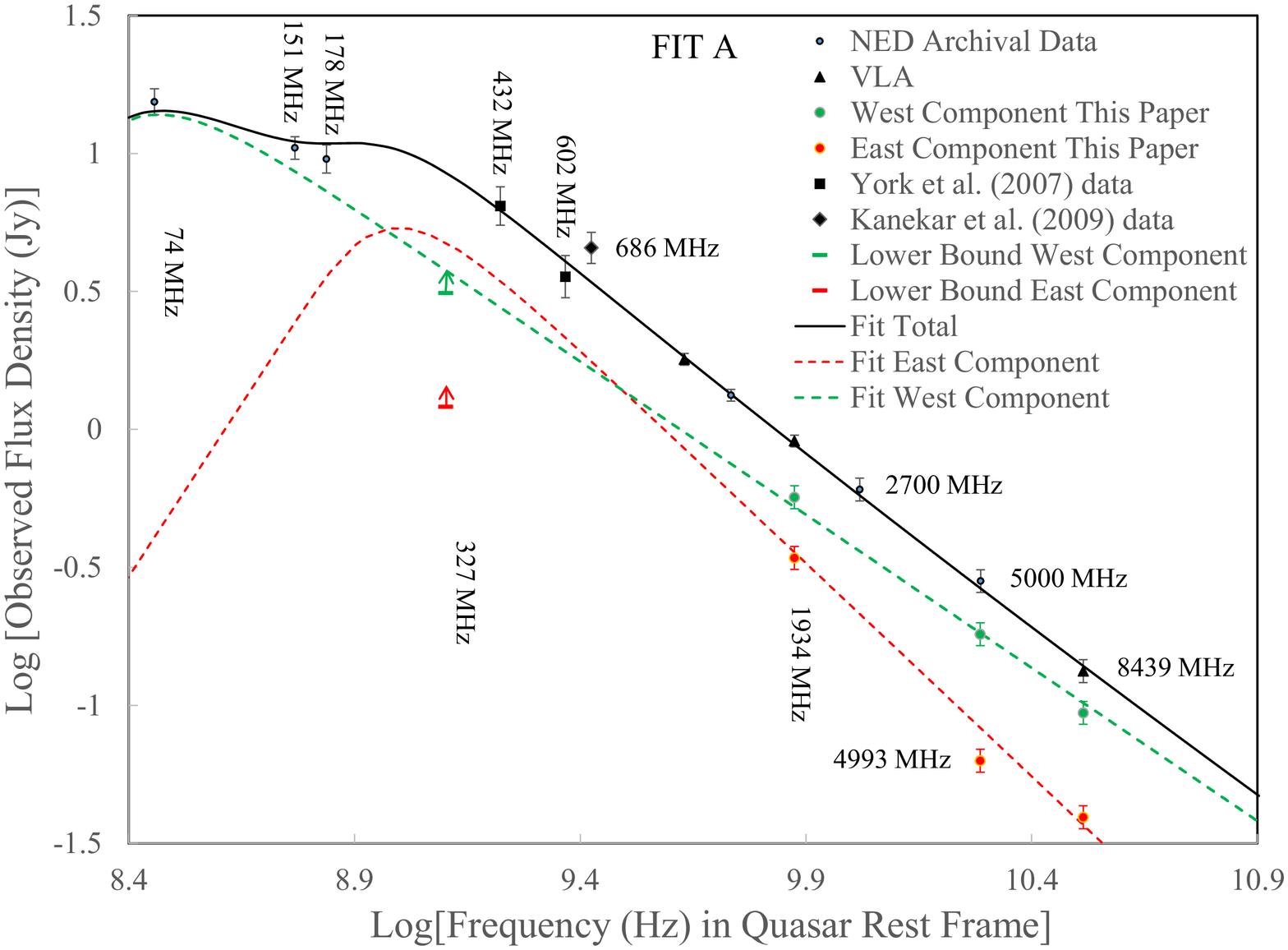}

\caption{The broad band radio spectrum of 3C 82 is well sampled by archival data from NED, our VLA data reductions, GBT and GMRT observations. The component flux densities are based on the data reductions presented here. Note that the MERLIN $\nu_{o}=5$ GHz and the VLBA $\nu_{o}=327$ MHz component flux densities seems to miss considerable diffuse emission in the eastern lobe. Being at such high resolution, with patchy u-v coverage, the $\nu_{o}=327$ MHz data points should formally be considered as lower bounds to the flux density of the diffuse lobes (see Table 2). Observed frequencies of data points are labeled in MHz. We overlay ``Fit A" that is described in Table 3 and Sections 5 and 6. }
\end{center}
\end{figure*}
\begin{table}
    \caption{Radio Data for 3C82}
{\tiny\begin{tabular}{cccccc} \tableline\rule{0mm}{3mm}
$\nu_{o}$ & $\log{\nu}$ &Flux & Telescope & Reference & Comments\\
Observed  &  Rest   & Density &  &  &  \\
Frequency   & Frame  &  &  &  &  \\
(MHz)   & (Hz)  & (Jy) & &  &  \\
\tableline \rule{0mm}{3mm}
73.8 & 8.46 & $15.390\pm 1.539$ & VLA B-array & \citet{coh07}   &  \\
150 &    8.76      & $10.140\pm 1.521$ \tablenotemark{a} & GMRT & \citet{int17}  &  \\
151.5 &    8.77      & $10.470\pm 1.047$  & 6C Telescope & \citet{hal93}  &  \\
178 &    8.84      & $9.55\pm 1.07$\tablenotemark{b}  & Large Cambridge Interferometer & \citet{gow67}  &  \\
432  &   9.22     & $6.450\pm 0.950$  & GBT & \citet{yor07}  &  \\
602 & 9.37 & $3.580\pm 0.587$  & GMRT & \citet{yor07} &   \\
686 & 9.42 & $4.550\pm 0.550$  & GBT & \citet{kan09} &   \\
1104 & 9.63 & $1.794\pm 0.090$  & JVLA A-array & this paper &   \\
1400 & 9.73 & $1.330\pm 0.067$  & VLA D-array & \citet{con98} &   \\
1934 & 9.87 & $0.908\pm 0.045$  & JVLA A-array & this paper &   \\
2700 & 10.02 & $0.606\pm 0.061$  & One-Mile Telescope & \citet{ril89} &   \\
5000 & 10.29 & $0.282\pm 0.028$  & One-Mile Telescope & \citet{ril89} &   \\
8439 & 10.51 & $0.133\pm 0.013$  & VLA A-array & this paper &   \\
\tableline \rule{0mm}{3mm}
East Lobe &  &  &  &  &  \\
327 & 9.10 & 1.21, Lower Bound  & VLBA & \citet{kan13} & Over-Resolved  \\
1934 & 9.87 & $0.342\pm 0.034$  & JVLA A-array & this paper &   \\
4993 & 10.29 & $0.063\pm 0.006$  & MERLIN & this paper & Over-Resolved  \\
8439 & 10.51 & $0.039\pm 0.004$  & VLA A-array & this paper &   \\
\tableline \rule{0mm}{3mm}
West Lobe &  &  &  &  &  \\
 327 & 9.10 & 3.12, Lower Bound  & VLBA & \citet{kan13} & Over-Resolved  \\
1934 & 9.87 & $0.568\pm 0.057$  & JVLA A-array & this paper &   \\
4993 & 10.29 & $0.181\pm 0.018$  & MERLIN & this paper & Over-Resolved  \\
8439 & 10.51 & $0.094\pm 0.009$  & VLA A-array & this paper &
\end{tabular}}
\tablenotetext{a}{Uncertainty based on considerations of \citet{hur17}.}
\tablenotetext{b}{Rescaling of \citet{gow67} to the \citet{baa77} scale by \citet{kuh81}.}
\end{table}
\par There is no clear detection of a compact radio core in 3C82. We looked at the optical position in order to see if it can provide some information on the location of the central engine and radio core. Without HST imaging combined with astrometrically accurate radio images, it is difficult to tie radio and optical positions together at a level where the optical position determines a precise physical location in such a very small radio structure. We used the Pan-STARRS1 survey, data release 2, since this has the highest astrometric accuracy of any ground based optical survey \footnote{https://catalogs.mast.stsci.edu/panstarrs/} \citep{cha19}. The optical position (RA = 3h14m43.6624s, Dec = +43d14m05.1761s) is overlaid on the images in Figure 3 as a black cross representing the error bars on position. The uncertainty is the standard deviation of the mean of the 61 position measurements made during the survey ($\sigma_{RA} = 0.111$ arcsec, $\sigma_{DEC} = 0.127$ arcsec). However, the radio images were created with phase self-calibration so they can be prone to larger positional errors. Even when phase referencing is employed, significant positional offsets can be induced by the residual phase in the transfer of the interpolated phases to the target field. If the radio observations lack astrometric accuracy, the location of the central engine of the radio source using the optical position is poorly constrained. The MERLIN position is the most accurate. It is phase referenced with a nearby compact phase calibrator (4 degrees from the target). The positional accuracy is $\sim 8$ mas for our observation\footnote{The calculation was performed by Anita Richards of the MERLIN/VLBI National Facility}. So, if the two VLA 8.4 GHz flux density peaks (one for the east lobe and one for the west lobe) are chosen to align with the two MERLIN 5 GHz flux density peaks, the resultant coordinate shift moves the X-band image (north by 0.24 arcsec and west by 0.19 arcsec) relative to the optically determined (therefore stationary) Pan-STARRS1 position. It is the relative position of the radio source and the optical position that is relevant to the physical nature of 3C82. In order to illustrate the Pan-STARRS1 position relative to the details of the X-band image with maximum astrometrical accuracy, we shift the Pan-STARRS1 position by opposite of this amount (south by 0.24 arcsec and east by 0.19 arcsec) on the background of the radio image.in the top left hand panel of Figure 3. The black cross is the position assuming the X-band astrometry is correct and the red cross to the south east is corrected for the reliable astrometry from the Merlin image. This adjusted optical position is the appropriate one to use in examining the relationship between the X-band radio structure and the optical position. The optical position is $< 100$~mas (i.e., within the error bars) from the tip of putative jet. We can only speculate the following about the faint radio core:
\begin{itemize}
\item The radio core might be heavily synchrotron self absorbed at these frequencies.
\item The radio core might be located at the tip of the putative jet in the X-band image. The hypothesized configuration would be very similar to B3 0710+439, a compact symmetric object. The central engine of this quasar has been identified with the low flux density tip of a jet \citep{rea96}.
\end{itemize}

\par Figure 4 and Table 2 present the radio data used in the physical analysis that follows. The spectral data is plotted in terms of quasar rest frame frequencies (see Table 2), since these are  the relevant frequencies required for understanding the physical source of the radio emission. It will be necessary to understand the discussion in terms of both the rest frame frequency, $\nu$, (for physical context) and the frequency, $\nu_{o}$, of the observations. Table 2 can help with this cross referencing. Fit A is described in the discussion of Table 3 and in Sections 5 and 6. It is superimposed on the data. We need to include archival data in order to get the low frequency spectrum. When multiple observations exist in the archives at the same frequency, we choose the data with the smallest field of view in order to avoid source confusion, the main source of flux density errors for this bright small source. We use survey data sparingly in our analysis, except at low frequency. Our most reliable survey point is the $\nu_{o}=151$ MHz ($\nu=584$ MHz in the quasar rest frame) observation from the 6C survey. There is the possibility of source confusion in such a wide field of view. We validated the observation by comparing the measured 10.47 Jy to the GMRT 150 MHz survey data in the TIFR GMRT Sky Survey Alternative Data Release (TGSS ADR), 10.14 Jy \citep{int17} \footnote{https://vo.astron.nl/tgssadr/q/cone/form}. We prefer the 6C data because in our experience the 10\% uncertainty used in TGSS ADR does not seem to be vetted well on a case by case basis and can be considerably larger for individual sources and 15\% uncertainty is a more prudent choice \citep{hur17}. This redundant data does not appear in our plots. We also have used the much older $\nu_{o}=178$ MHz ($\nu=689$ MHz in the quasar rest frame) 3C survey data in order to improve coverage in this region \citep{gow67}. The $\nu_{o}=432$ MHz ($\nu=1.67$ GHz in the quasar rest frame) and the $\nu_{o}=151$ MHz data are important in conjunction with the $\nu_{o}=74$ MHz ($\nu=287$ MHz in the quasar rest frame) as they are the only data that lie on what appears to be a very broad spectral peak. The relatively ``flat spectrum" in this region appears as a pronounced break in the apparent very steep power law spectrum (spectral index, $\alpha \approx 1.2$) that extends from $\nu_{o}=432$ MHz to $\nu_{o}=8.44$ GHz ($\nu = 32.66$ GHz in the quasar rest frame). It is this steep power law and the broad low frequency flat spectrum region defined by the radio data that will constrain the physical models of the radio lobes in Sections 5 and 6.

\section{Synchrotron Self-Absorbed Homogeneous Plasmoids}
Based on the images in Figure 3, the radio flux of 3C 82 is dominated by the two radio lobes. Complicated time evolving dynamics have been inferred to occur in radio lobes \citep{blu00}. In general there are fine-scale features like hot spots, portions of jets and filaments embedded within a diffuse lobe plasma. However, our images show very little structure, perhaps a weak hot spot in the VLBA $\nu_{o}=327$ MHz image, and perhaps a weak jet at $\nu_{o}=8.4$ GHz. To an excellent approximation the vast majority of the emission from $\nu_{o}=1.9$ GHz to $\nu_{o}=8.4$ GHz is located in two unresolved blobs. The Gaussian fitting procedure indicates that there is not enough information to reliably decompose the lobes into diffuse steep spectrum lobe plasma, and finer scale, flatter spectrum features \citep{liu92,fer07,fer14}. Thus, we use simple homogeneous, uniform single zone models of plasma in order to approximate the physics. These single zone spherical models are even a standard technique in blazar jet calculations out of practical necessity \citep{ghi10}. Thus motivated, we describe a simple two uniform spherical ``plasmoid" model, one uniform spherical zone per radio lobe. The difference between a uniform spherical zone and a spherical plasmoid is merely semantics. The simple homogeneous spherical volume model has historically provided an understanding of the spectra and the time evolution of astrophysical radio sources \citep{van66}. The specific application of this model to be implemented here has been used to study a variety of problems such as the major flares in the Galactic black hole accretion system of GRS~1915+105, (Punsly, 2012), the neutron star binary merger that was the gravity wave source, GW170817, and associated gamma ray burst, GRB 170817A, (Punsly, 2019), and radio flares in the quasar Mrk~231 \citep{rey09,rey20}. The primary advantage of the method is that the SSA turnover provides information on the size of the region that produces the preponderance of emission. This cannot be obtained with adequate accuracy from images with insufficient resolution and sensitivity. For example, assuming a size equal to the resolution limit of the telescope generally results in plasmoid energy estimates that are off by one or more orders of magnitude due to the exaggerated volume of plasma \citep{fen99,pun12}. Figure 4 seems to indicate two relative maxima in the radio spectrum, one near $\nu=1$ GHz in the quasar rest frame and one near $\nu=300$ MHz. We use these peaks to set the magnitude of the SSA. The first subsection will describe the underlying physics and the next subsection describes the mechanical energy flux in the spherical plasmoids.

\subsection{The Underlying Physical Equations}
One must differentiate between quantities measured in the plasmoid frame of reference and those measured in the observer's frame of reference. The physics is evaluated in the plasma rest frame. Then, the results are transformed to the observer's frame for comparison with observation. The underlying powerlaw for the flux density is defined as $S_{\nu}(\nu= \nu_{o}) = S\nu_{o}^{-\alpha}$, where $S$ is a constant. Observed quantities will
be designated with a subscript, ``o", in the following expressions. The observed frequency is related to the emitted frequency, $\nu_{e}$, by $\nu_{o} = \delta \nu_{e}/(1+z)$. The bulk flow Doppler factor of the plasmoid, $\delta$,
\begin{equation}
\delta = \frac{\gamma^{-1}}{1-\beta \cos{\theta}},\; \gamma^{-2} = 1- \beta^{2}\;,
\end{equation}
where $\beta$ is the normalized three-velocity of bulk motion and $\theta$ is the angle of the motion to the line of sight (LOS) to the observer.
The SSA attenuation coefficient is computed in the plasma rest frame \citep{gin69},
\begin{eqnarray}
&& \mu(\nu_{e})=\overline{g(n)}\frac{e^{3}}{2\pi
m_{e}}N_{\Gamma}(m_{e}c^{2})^{2\alpha} \left(\frac{3e}{2\pi
m_{e}^{3} c^{5}}\right)^{\frac{1+2\alpha}{2}}\left(B\right)^{(1.5
+\alpha)}\left(\nu_{e}\right)^{-(2.5 + \alpha)}\;,\\
&& \overline{g(n)}= \frac{\sqrt{3\pi}}{8}\frac{\overline{\Gamma}[(3n
+ 22)/12]\overline{\Gamma}[(3n + 2)/12]\overline{\Gamma}[(n +
6)/4]}{\overline{\Gamma}[(n + 8)/4]}\;, \\
&& N=\int_{\Gamma_{min}}^{\Gamma_{max}}{N_{\Gamma}\Gamma^{-n}\,
d\Gamma}\;,\; n= 2\alpha +1 \;,
\end{eqnarray}
where $\Gamma$ is the ratio of lepton energy to rest mass energy, $m_{e}c^2$, $\overline{g(n)}$ is the Gaunt factor averaged over angle and $\overline{\Gamma}$ is the gamma function. $B$ is the magnitude of the total
magnetic field. The low energy cutoff, $E_{min} = \Gamma_{min}m_{e}c^2$, is constrained loosely by the data in Figure 4. The fact that 3C82 is very luminous at
 $\nu =200 $~MHz, means that the lepton energy distribution is not cut off near this frequency. This is not a very stringent bound. Note that the SSA opacity in the observer's frame, $\mu(\nu_{o})$, is obtained by substituting $\nu_{e} =(1+z)\nu_{o} / \delta$ into Equation (2). In the following analysis $\delta\approx 1$, so $\nu_{e} \approx \nu =(1+z)\nu_{o}$. So we will drop the subscript, ``e" in order to streamline the notation i.e., the plasmoid frame of reference and the quasar cosmological frame of reference are the same up to negligible relativistic corrections.

\par A simple solution to the radiative transfer equation occurs in the homogeneous approximation \citep{gin65,van66}
\begin{eqnarray}
&& S_{\nu_{\o}} = \frac{S_{o}\nu_{o}^{-\alpha}}{\tau(\nu_{o})} \times \left(1 -e^{-\tau(\nu_{o})}\right)\;, \; \tau(\nu_{o}) \equiv \mu(\nu_{o}) L\;, \; \tau(\nu_{o})=\overline{\tau}\nu_{o}^{(-2.5 +\alpha)}\;,
\end{eqnarray}
where $\tau(\nu)$ is the SSA opacity, $L$ is the path length in the rest frame of the plasma, $S_{o}$ is a normalization factor and $\overline{\tau}$ is a constant. There are three unknowns in Equation (7), $\overline{\tau}$, $\alpha$ and $S_{o}$. These are effectively three constraints on the following theoretical model that can be estimated from the observational data. These three constraints are used to establish the uniqueness and the existence of physical solutions in Sections 5 and 6.
\par In order to connect the parametric spectrum given by Equation (7) to a physical model requires an expression for the synchrotron emissivity \citep{tuc75}:
\begin{eqnarray}
&& j_{\nu} = 1.7 \times 10^{-21} [4 \pi N_{\Gamma}]a(n)B^{(1
+\alpha)}\left(\frac{4
\times 10^{6}}{\nu}\right)^{\alpha}\;,\\
&& a(n)=\frac{\left(2^{\frac{n-1}{2}}\sqrt{3}\right)
\overline{\Gamma}\left(\frac{3n-1}{12}\right)\overline{\Gamma}\left(\frac{3n+19}{12}\right)
\overline{\Gamma}\left(\frac{n+5}{4}\right)}
       {8\sqrt\pi(n+1)\overline{\Gamma}\left(\frac{n+7}{4}\right)} \;,
\end{eqnarray}
where the coefficient $a(n)$ separates the pure dependence on $n$ \citep{gin65}.
One can transform this to the observed flux density, $S(\nu_{o})$, in the optically thin region of the spectrum (i.e., our VLA and JVLA measurements in the case of 3C82) using the relativistic transformation relations from \citet{lin85},
\begin{eqnarray}
 && S(\nu_{o}) = \frac{\delta^{(3 + \alpha)}}{4\pi D_{L}^{2}}\int{j_{\nu}^{'} d V{'}}\;,
\end{eqnarray}
where $D_{L}$ is the luminosity distance and in this expression, the
primed frame is the rest frame of the plasma. Equations (4) - (10) are used in Section 5 to fit the data in Figure 4.

\subsection{Mechanical Contributions to the Energy Flux} The energy content is separated into two parts. The first is the kinetic energy of the protons, $KE(\mathrm{proton})$,
\begin{eqnarray}
 && KE(\mathrm{protonic}) = (\gamma - 1)Mc^{2}\;,
\end{eqnarray}
here $M$ is the mass of the plasmoid. The other piece is the lepto-magnetic energy, $E(\mathrm{lm})$, and is composed of the volume integral of the leptonic internal energy density, $U_{e}$, and the magnetic field energy density, $U_{B}$. The lepto-magnetic energy in a uniform spherical volume is
\begin{eqnarray}
 && E(\mathrm{lm}) = \int{(U_{B}+ U_{e})}\, dV = \frac{4}{3}\pi R^{3}\left[\frac{B^{2}}{8\pi}
+ \int_{\Gamma_{min}}^{\Gamma_{max}}(m_{e}c^{2})(N_{\Gamma}E^{-n + 1})\, d\,E \right]\;.
\end{eqnarray}
It has been argued that the lepto-magnetic energy is often nearly minimized in the hot spots and enveloping lobes of large FRII radio sources \citep{har04,cro05,kat05}. However, as we discuss in section 7.3, this condition must be evaluated on a case by case basis. The leptons also have a kinetic energy analogous to Equation (11),
\begin{eqnarray}
 && KE(\mathrm{leptonic}) = (\gamma - 1)\mathcal{N}_{e}m_{e}c^{2}\;,
\end{eqnarray}
where $\mathcal{N}_{e}$ is the total number of leptons in the plasmoid.
\begin{figure*}
\begin{center}
\includegraphics[width= 0.5\textwidth,angle =0]{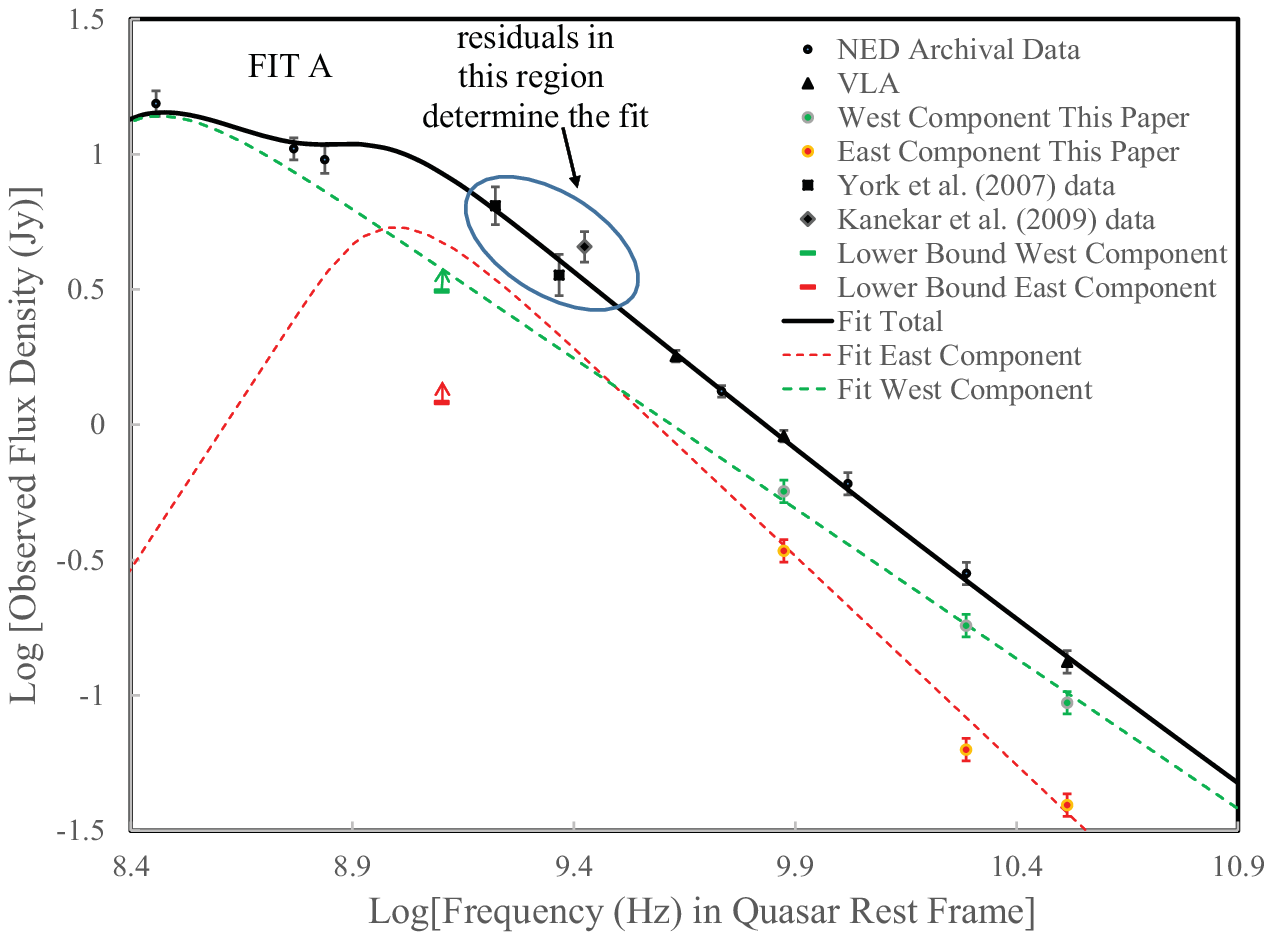}
\includegraphics[width= 0.5\textwidth,angle =0]{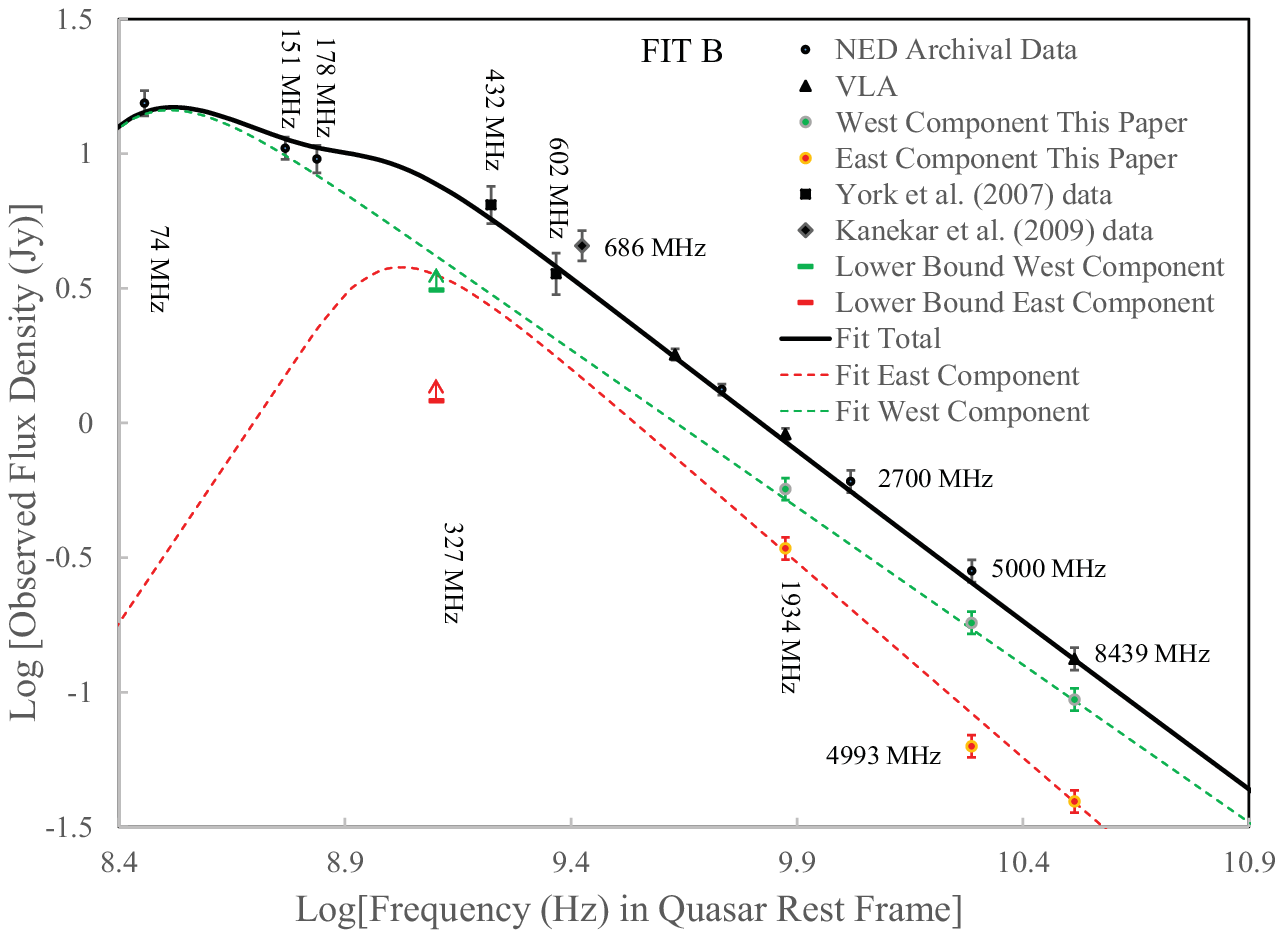}
\includegraphics[width= 0.5\textwidth,angle =0]{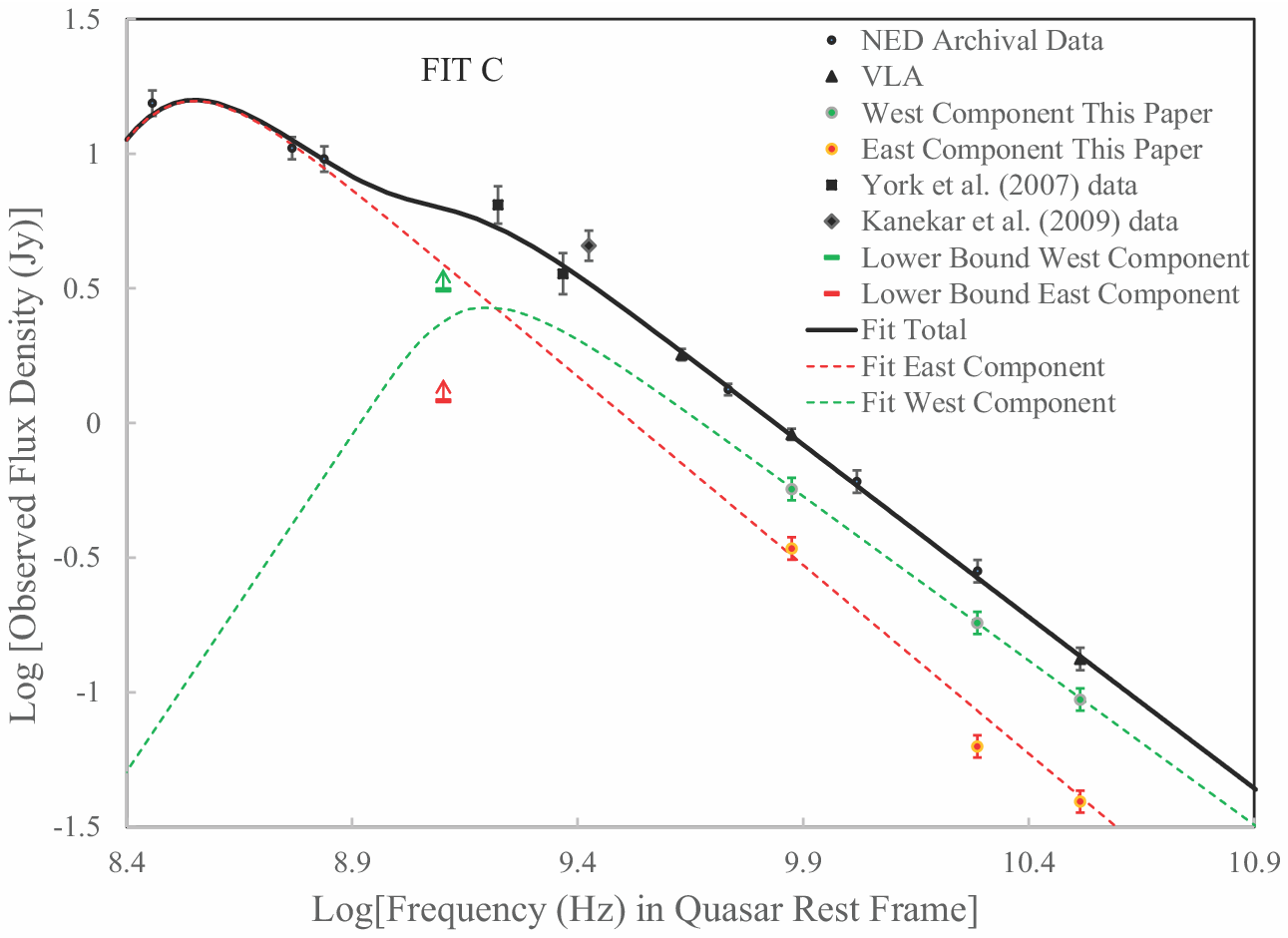}
\caption{\footnotesize{A comparison of the three fits described in Sections 6.1-6.3 and Table 3. Fit A is repeated from Figure 4, with one modification, we place a blue oval around the three data points at the low frequency end of the power law region of the spectrum, adjacent to the low frequency turnover. The tension between the low frequency turnover and the adjacent power law region makes for a tight fit as discussed in the text. This region also generates the largest residuals in the fitting process. Fit A has the smallest residuals of the fits presented.}}
\end{center}
\end{figure*}

\section{Fitting the Data with a Specific Spherical Model}
Inspection of the radio data in Figure 4, shows that 3C82 is well described by a power law from $\nu=32.67$ GHz to about $\nu=1.67$ GHz with $\alpha \approx 1.2$. But, the spectrum is not just a power law. The spectrum turns over and flattens towards low frequency. The turnover is not a simple shape with a monotonically changing curvature. It appears to have two relative maxima as defined by the sparse data at $\nu_{o}=74$ MHz, $\nu_{o} =151$ MHz, $\nu_{o} =178$ MHz and $\nu_{o}=432$ MHz. There is a relative maximum near $\nu=900$ MHz (between the $\nu_{o}=151$ MHz and $\nu_{o}=432$ MHz data). There is a second relative maximum near $\nu=300$ MHz (near the $\nu_{o}=74$ MHz data). The spectrum can be described by 4 observationally determined parameters. Two derive from the power law $\alpha$, $S_{o}$. Two others are from the two relative maxima in the SSA region. In practice, this means that the two parameter power law above $\nu_{o}=432$ MHz must be fit simultaneously with the $\nu_{o}=74$ MHz, the $\nu_{o} =151$ MHz and the $\nu_{o} =178$ MHz data points. It is the tension between fitting these data simultaneously that makes the fit very tight. This complex spectrum cannot be fit with a single SSA power law as defined in Equation (7).

\par The double humped nature of the SSA spectral peak suggests a decomposition into two SSA power laws, one for each lobe. From Equation (7) each SSA power law has 3 parameters, $\overline{\tau}$, $\alpha$, $S_{o}$. For the two lobes this will be 6 parameters: $\overline{\tau}(west)$, $\overline{\tau}(east)$, $\alpha(west)$, $\alpha(east)$, $S_{o}(west)$ and $S_{o}(east)$. Ostensibly, this decomposition uses 6 parameters to determine 4 observational constraints and is under-determined. But, there is degeneracy in the constraints from the radio data for which we have derived component flux densities, so this is really not the case. Namely, by extracting component flux densities from our radio images, we know more than just the total flux density. The Gaussian fit models determine $\alpha(west)$, $\alpha(east)$, $S_{o}(west)$ and $S_{o}(east)$. These component values will produce $\alpha$ and $S_{o}$ in the combined spectrum up to the uncertainty in the error bars. The fitting procedure is then reduced to two unknowns and two constraints. We adjust $\overline{\tau}(west)$, $\overline{\tau}(east)$ in order to fit the $\nu_{o}=151$ MHz and the $\nu_{o}=74$ MHz data within the error bars. This choice results in a small feedback on the power law i.e., this requires minor adjustments to the component power laws). The power law adjustments are restricted by the error bars on the total flux density measurements at frequencies above $\nu_{o}=432$ MHz and the error bars on the flux densities of the Gaussian two component fits. Thus, the fit is not unique, but there is not much variation allowed by the radio data. The only major uncertainty is which lobe is most responsible for the $\nu_{o}=74$ MHz flux density, but in Section 5.3, we show that the data determines this as well.
\par Once an SSA power law is chosen for each lobe, we note that from a mathematical perspective, the theoretical determination of $S_{\nu}$ depends on 7 physical parameters in Equations (4)-(10), $N_{\Gamma}$, $B$, $R$ (the radius of the sphere), $\alpha$, $\delta$, $E_{min}$ and $E_{max}$, yet there are only 3 constraints from the SSA power law model, $\overline{\tau}$, $\alpha$ and $S_{o}$. This is an under-determined system of equations. Firstly, in Section 3, we found that the radio lobes are very steep spectrum above $\nu=1.67$ GHz in the quasar rest frame. Therefore, most of the leptons are at low energy, and the solutions are insensitive to $E_{max}$.

\subsection{The Lobe Doppler Factor} In terms of estimating the relevant Doppler factor for the lobes, we have no detection of lobe plasma motion in any radio lobe 10kpc from the central engine in any AGN. We must rely on theoretical arguments such as those based on synchrotron cooling and the spatial evolution of spectral breaks across the radio lobe \citep{liu92,ale96}. The velocity of the diffuse lobe plasma, $v_{diffuse}$, is a superposition of hotspot advance speed, $v_{hs}$, the back flow speed, $v_{bf}$, and the lateral expansion speed, $v_{exp}$. It has been deduced that $v_{hs}$ and $v_{bf}$ are the largest contributors and $v_{hs} \approx v_{bf}$, almost cancelling \citep{ale96,liu92}. The instantaneous $v_{hs}$ is a different quantity than the lobe advance speed averaged over the entire jet history, $v_{adv}$, used in statistical studies \citep{sch95}. Estimating $v_{hs}$ at a given time in a given radio source's history is difficult and necessarily speculative. In order to make an estimate of $v_{hs}$ requires high resolution images at multiple frequencies in order to deduce the synchrotron cooling as plasma flows away from the hot spot. It also requires high resolution X-ray observations that can be used to estimate the enveloping density and temperature. These elements are available for Cygnus A for which $v_{hs}\approx 0.005 c$ has been estimated \citep{ale96}. This best understood example can help to constrain 3C 82 which has none of the relevant information. $v_{hs}$ decelerates as the jet propagates \citep{ale06}. Cygnus A is an order of magnitude larger than 3C 82, so we expect $v_{hs}$ to be larger in 3C82. It is expected that $v_{hs}$ is at least factor of a few larger than 0.005c and significantly less than $v_{adv}$ estimated in Section 7. So, we crudely guess $0.05c<v_{hs}<0.1c$, with the condition $v_{diffuse} \ll v_{hs}$ \citep{ale96}. Since the rest frame UV is only mildly variable and the radio core is very weak, we assume a typical non-blazar line of sight of $30^{\circ}$ \citep{bar89}. Using the expression for $\delta$ in Equation (1), the ratio of Doppler enhancement of the approaching hot spot to the receding hot spot from Equation (10) is $1.4 < [\delta(\beta=v_{hs}/c)/\delta(\beta=-v_{hs}/c)]^{4}<2$. By contrast, $v_{diffuse}$ is a superposition of cancelling sub-relativistic velocities and $[\delta(\beta = v_{diffuse}/c)/\delta(\beta= -v_{diffuse}/c)]^{4}\approx 1$. If we could segregate the hot spot flux density with very high resolution, high sensitivity images, the Doppler enhancement would lower the energy in the our model of the approaching lobe (since the intrinsic luminosity is less than observed) and viceversa for the receding lobe (this will be shown explicitly in Section 6.5 for our models). But we do not have these images and we have no basis to make this decomposition. So, initially we choose $\delta = 1$. Then in section 6.5, we consider two cases in which the entire approaching (receding) lobe has a single velocity, $v_{lobe}=0.05c$ ($v_{lobe} = -0.05c$) and $v_{lobe}=0.1c$ ($v_{lobe} = -0.1c$) in order to assess the effects. The total energy of the system is unchanged under this range of properties.

\subsection{Solving the Reduced Set of Equations} Setting $\delta=1$ and treating $E_{max}$ as basically infinite, effectively adds 2 more known quantities, making the situation 5 unknowns with 3 constraints. The system is still under-determined and we would like to improve this situation. In order to restrict the size of the solution space we need to constrain $E_{min}$. We choose $E_{min}\approx m_{e}c^{2}$. This assumption needs to be checked on a case by case basis for internal consistency. We will show in Section 6.7 that this is the preferred value of $E_{min}$ for various physical reasons in our models of 3C 82.
\begin{table}
\caption{SSA Powerlaw Fits to the Radio Lobes and Details of the Corresponding Model}
{\tiny\begin{tabular}{cccccccccc} \tableline\rule{0mm}{3mm}
(1) & (2)  & (3)  & (4)  & (5) & (6) & (7) & (8) & (9)& (10) \\
&  &   &  & & Excess & Excess& Minimum & Total& \\
 &   &  &  &  & Variance & Variance & Energy  & $E(lm)$ & \\
 &  &  &  &  & Powerlaw & Less & Solution & Eqn. 12 & Largest\\
  & &    &  &   & $\nu_{o}=$  & $\nu_{o}=686$ MHz  & $\frac{E(lm)_{\rm{east}}}{E(lm)_{\rm{west}}}$& ergs & Lobe\\
  & &    &  &   & 0.43-8.45 GHz & Outlier & &  & \\
  \hline
  &Region  & Peak  & Luminosity  & Spectral &  & &  & & \\
 &   & Frequency\tablenotemark{a} & at Spectral Peak & Index &  &  &   & & \\
 &  & (MHz) & (ergs/sec) & $\alpha$ & $\sigma_{\rm{rms}}^{2}$ & $\sigma_{\rm{rms}}^{2}$ &  &  & \\
  & &    $\nu_{\rm{peak}}$ & $\nu L_{\nu}(\nu = \nu_{\rm{peak}})$ &   &  &  & &  & \\
\tableline \rule{0mm}{3mm}
I. Fit A \tablenotemark{b} & Total & N/A  & N/A & N/A & +0.017& -0.011 & 0.77 &$1.55\times 10^{60}$ & West\\
\hline
Components &  &  &  &  & & &&&\\
\hline
& West Lobe  & 290 & $7.22 \times 10^{44}$ & 1.11  & N/A & N/A &N/A& N/A&N/A\\
& East Lobe & 950 & $9.48 \times 10^{44}$ & 1.55 &N/A &N/A &N/A&N/A&N/A\\
\hline
\hline
&  &  &  &  & & &&\\
\hline
II. Fit B\tablenotemark{c} & Total  & N/A & N/A  &  N/A & +0.042& -0.003 & 0.24&$1.37\times 10^{60}$& West\\
\hline
Components &  &  &  &  & & &&&\\
\hline
& West Lobe & 330 & $7.57 \times 10^{44}$ & 1.17  & N/A & N/A &N/A& N/A&N/A\\
& East Lobe & 1060 & $6.77 \times 10^{44}$ & 1.45 & N/A & N/A &N/A& N/A&N/A\\
\hline
\hline
&  &  &  &  & & &&&\\
\hline
III. Fit C\tablenotemark{d} & Total  & N/A &  N/A & N/A  & +0.050& +0.015 & 86& $3.58\times 10^{60}$& East\\
\hline
Components & &  &  &  & & &&&\\
\hline
& West Lobe & 1580 & $1.41 \times 10^{44}$ & 1.22  & N/A & N/A &N/A& N/A&N/A\\
& East Lobe & 260 & $2.80 \times 10^{45}$ & 1.40 & N/A & N/A &N/A& N/A&N/A\\
\hline
\hline
IV. Fit A & Total & N/A  & N/A & N/A & +0.017& -0.011 & 1.0 &$1.35 \times 10^{60}$ & West\\
Non-minimum\tablenotemark{e} &  &  &  &  & & &&&\\
Energy &  &  &  &  & & &&&\\
Section 6.4 &  &  &  &  & & &&&\\
\hline
Components &  &  &  &  & & &&&\\
\hline
& West Lobe  & 290 & $7.22 \times 10^{44}$ & 1.11  & N/A & N/A &N/A& N/A&N/A\\
& East Lobe & 950 & $9.48 \times 10^{44}$ & 1.55 &N/A &N/A &N/A&N/A&N/A\\
\hline
\hline
V. Fit A & Total & N/A  & N/A & N/A & +0.017 & -0.011 & 0.97 &$1.55 \times 10^{60}$ & West\\
$v_{lobe}=\pm 0.05c $ &  &  &  &  & & &&&\\
Section 6.5 &  &  &  &  & & &&&\\
\hline
Components &  &  &  &  & & &&&\\
\hline
& West Lobe  & 290 & $7.22 \times 10^{44}$ & 1.11  & N/A & N/A &N/A& N/A&N/A\\
& East Lobe & 950 & $9.48 \times 10^{44}$ & 1.55 &N/A &N/A &N/A&N/A&N/A\\
\hline
\hline
VI. Fit A & Total & N/A  & N/A & N/A & +0.017 & -0.011 & 1.26 &$1.54 \times 10^{60}$ & West\\
$v_{lobe}=\pm 0.1c $ &  &  &  &  & & &&&\\
Section 6.5 &  &  &  &  & & &&&\\
\hline
Components &  &  &  &  & & &&&\\
\hline
& West Lobe  & 290 & $7.22 \times 10^{44}$ & 1.11  & N/A & N/A &N/A& N/A&N/A\\
& East Lobe & 950 & $9.48 \times 10^{44}$ & 1.55 &N/A &N/A &N/A&N/A&N/A
\end{tabular}}
\tablenotetext{a}{Frequency in quasar rest frame}
\tablenotetext{b}{Maximum $\alpha$ in east lobe, minimum $\alpha$ in west lobe}
\tablenotetext{c}{Minimum $\alpha$ in east lobe, maximum $\alpha$ in west lobe}
\tablenotetext{d}{Frequencies of the spectral peaks of the lobes are switched. This is not a viable solution since the west component produces a 327 MHz flux density less than the VLBA lower bound.}
\tablenotetext{e}{Non-minimum model for west lobe with $E_{min} =5 m_{e}c^{2}$ }
\end{table}
\begin{figure*}
\begin{center}
\includegraphics[width= 0.45\textwidth,angle =0]{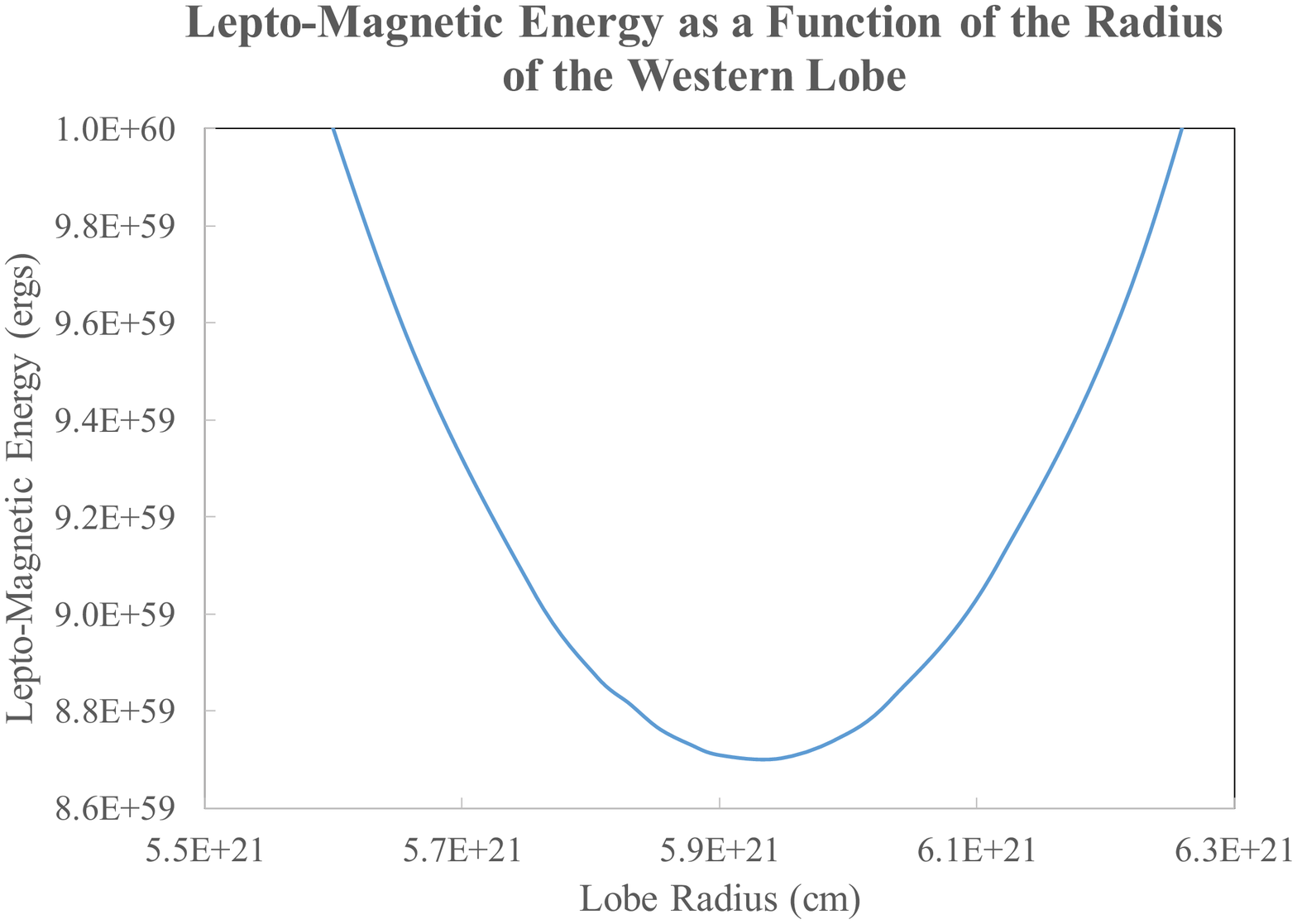}
\includegraphics[width= 0.45\textwidth,angle =0]{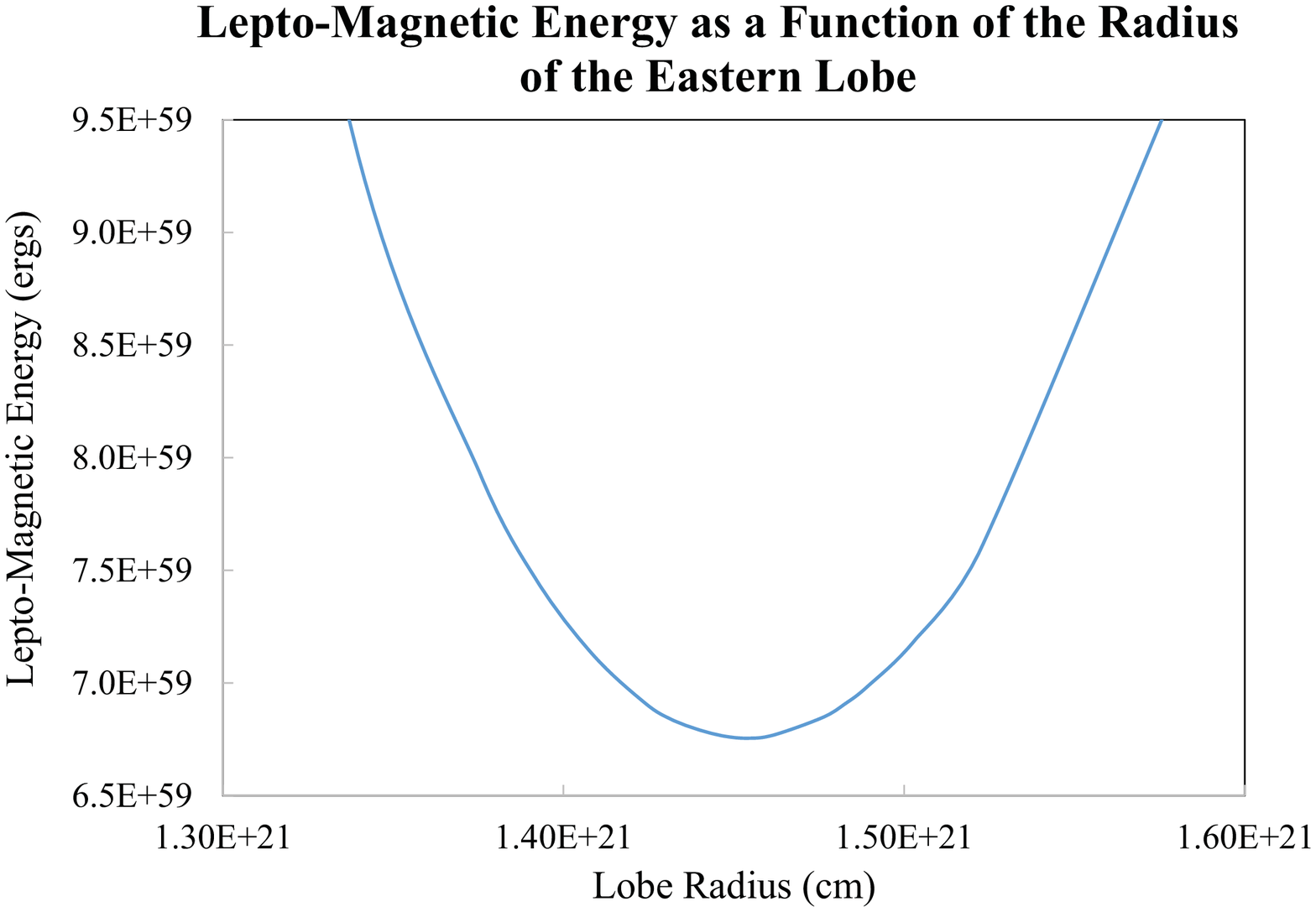}
\includegraphics[width= 0.45\textwidth,angle =0]{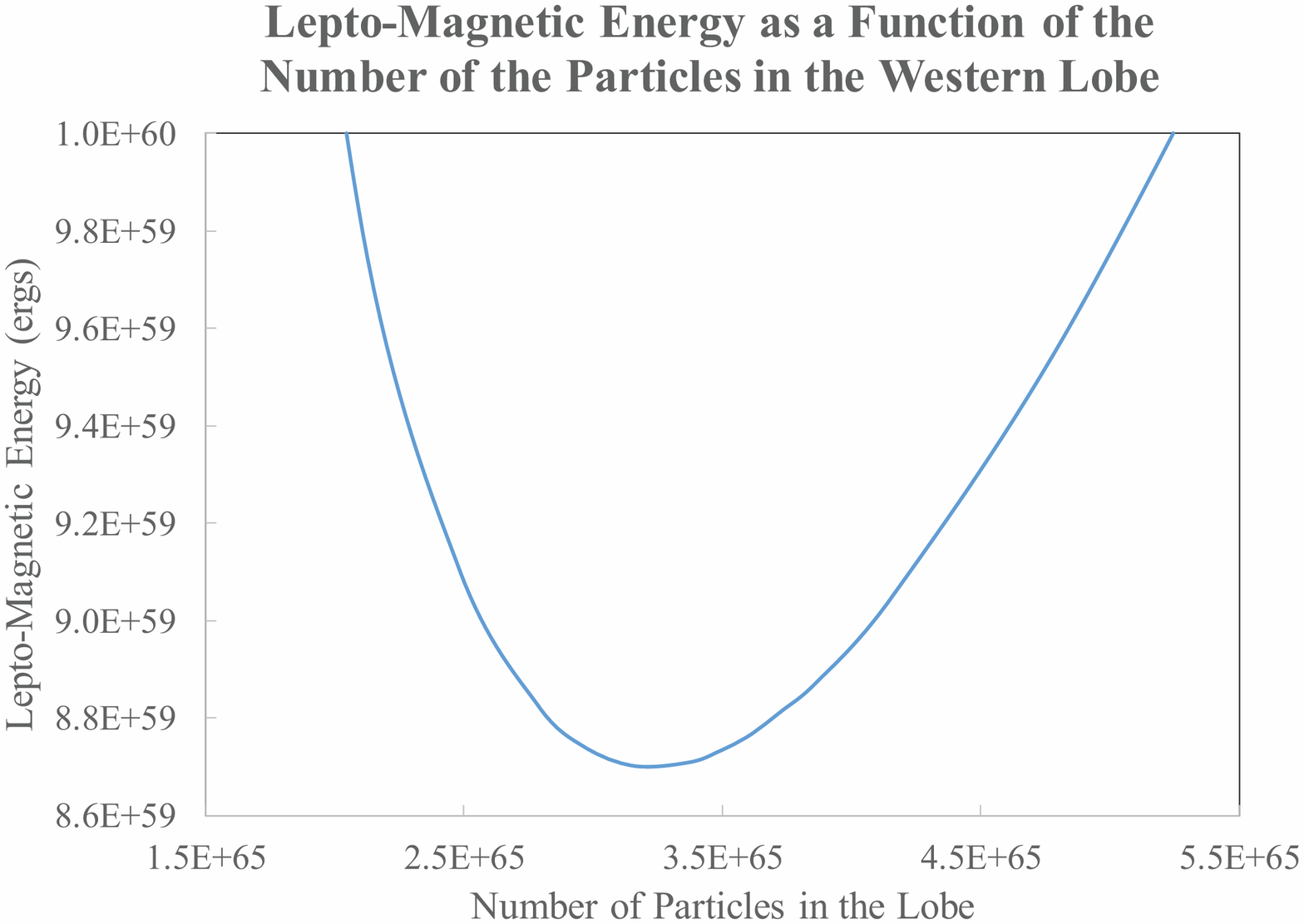}
\includegraphics[width= 0.45\textwidth,angle =0]{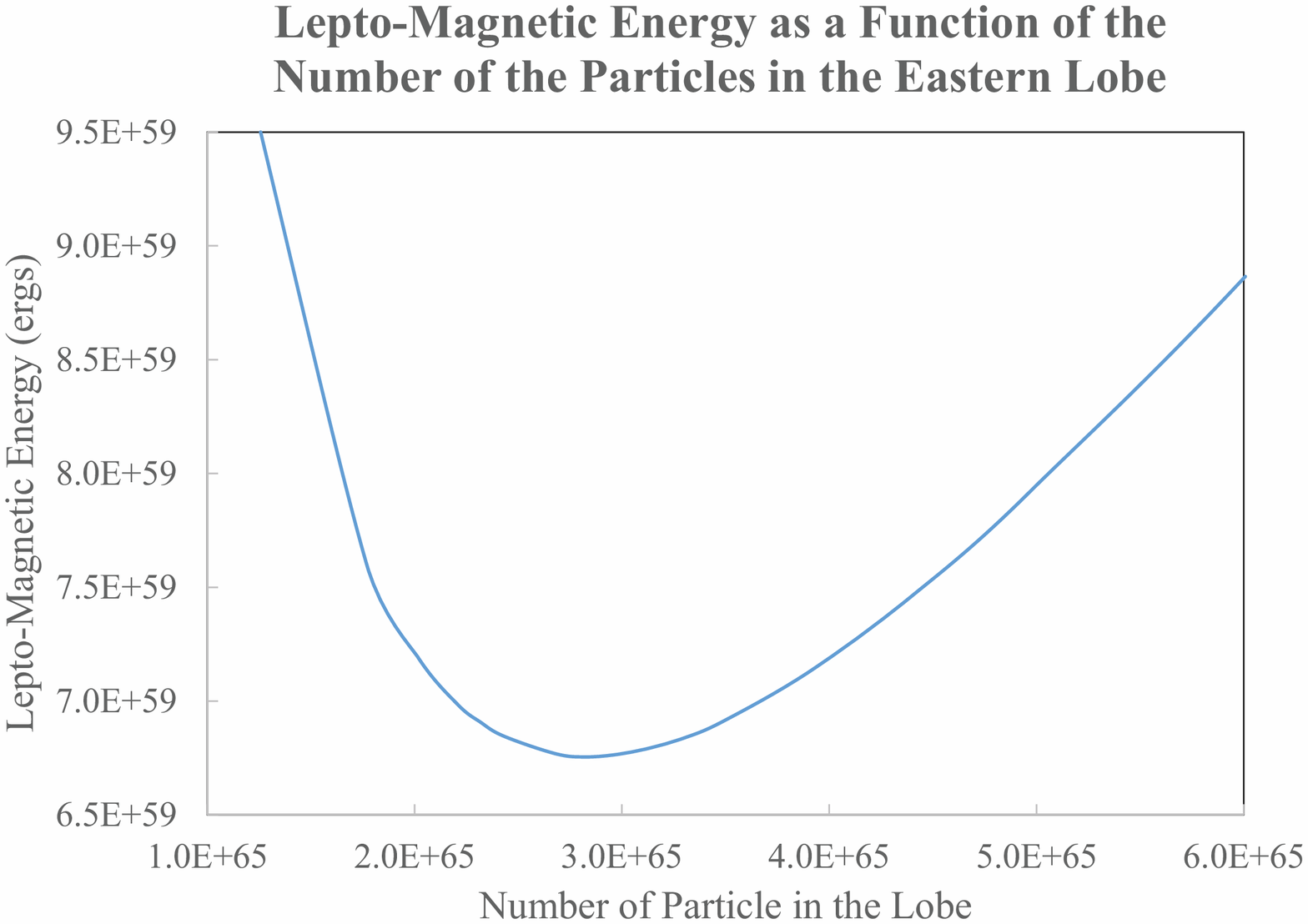}
\includegraphics[width= 0.45\textwidth,angle =0]{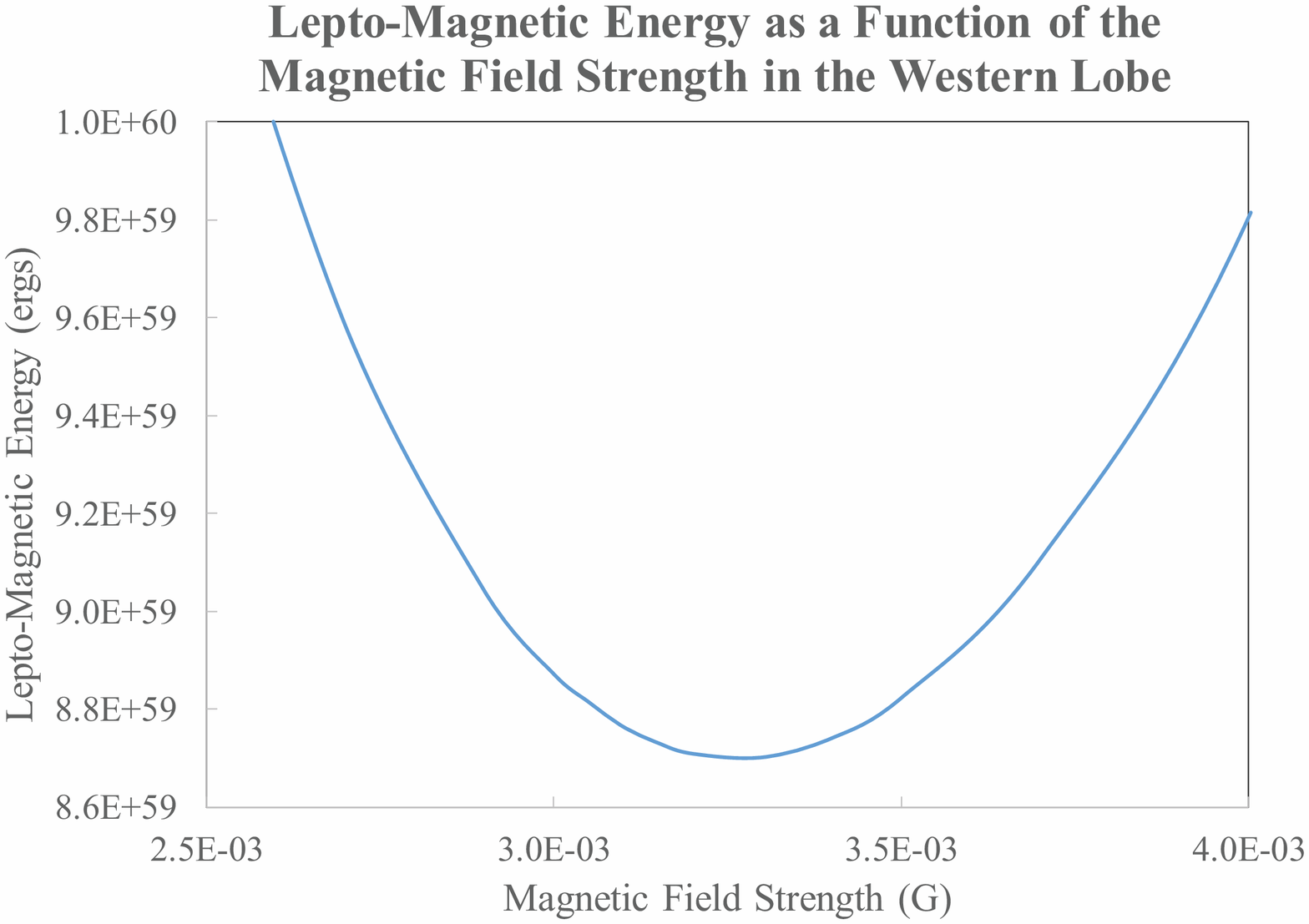}
\includegraphics[width= 0.45\textwidth,angle =0]{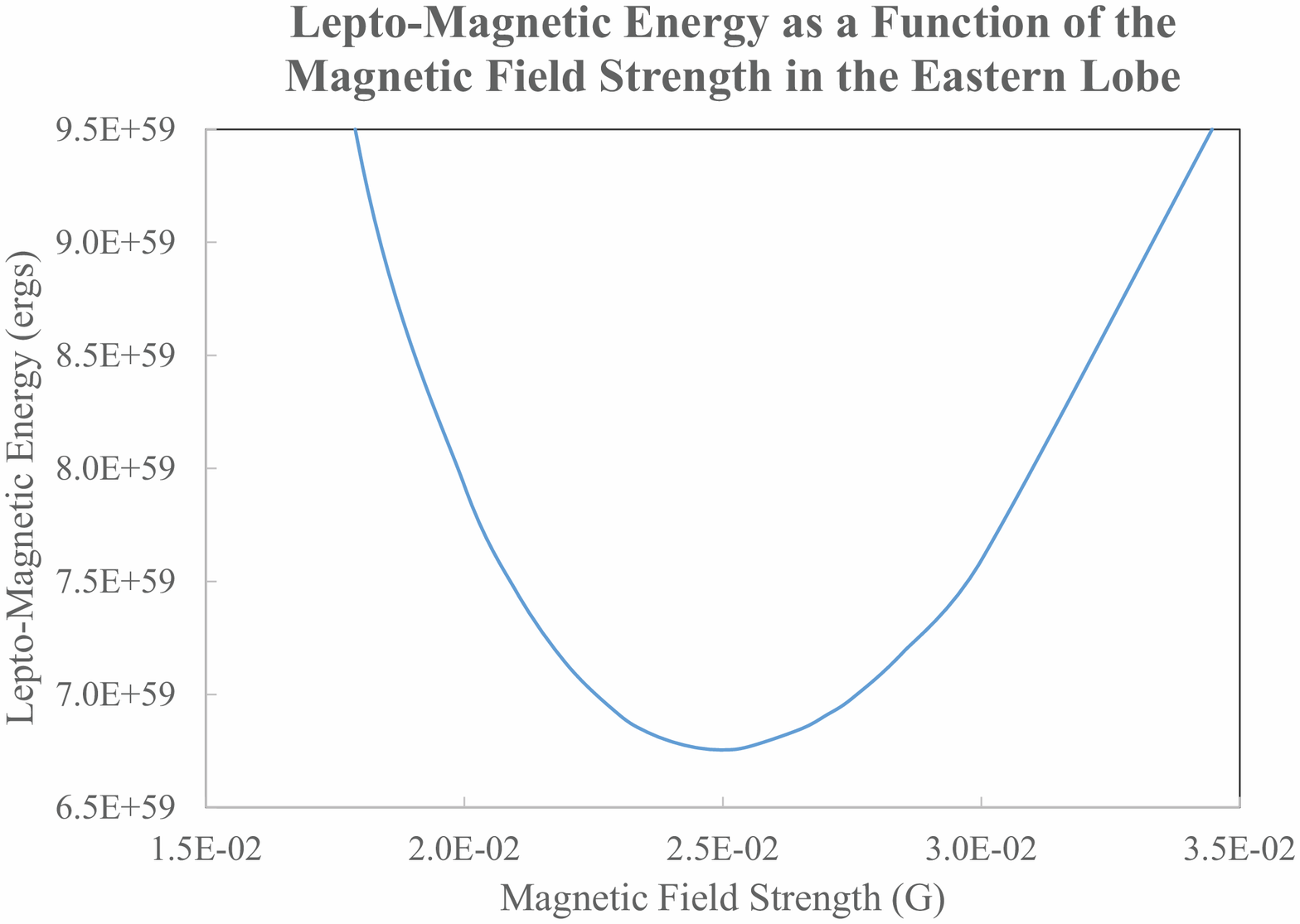}
\caption{The details of the minimum energy solution with $E_{min}=m_{e}c^{2}$ and $\delta=1$ for the fit to the radio spectrum in Figure 4, Fit A. The curves in these plots are the graphical manifestation of the infinite 1-D set of solutions for each lobe that was described in abstract mathematical terms in Section 5.2. The top two panels are the dependence of $E(\mathrm{lm})$ on the plasmoid radius, $R$, for the western lobe (left) and eastern lobe, (right). The next two rows are the dependence of $E(\mathrm{lm})$ on the total number of particles in the lobes and the magnetic field strength.}
\end{center}
\end{figure*}
\par  For the assumed values, $\delta=1$ and $E_{min}$, and recalling that the solutions are insensitive to $E_{max}$ one has 4 unknowns in each lobe, $N_{\Gamma}$, $B$, $R$, and $\alpha$. Yet each fitted SSA power law model has 3 fitted constraints, $\overline{\tau}$, $\alpha$, $S_{o}$. Thus, for each lobe there is an infinite 1 dimensional set of solutions for each pre-assigned $\delta$ and $E_{min}$ that results in the same spectral output. First, a power law fit to the high frequency optically thin synchrotron tail determined by our Gaussian component fits fixes the power law parameters in each lobe, $S_{o}$ and $\alpha$ in Equation (7). An arbitrary $B$ is chosen in each spheroid. Then $N_{\Gamma}$ and the spheroid radius, $R$, are iteratively varied to produce this fitted $S_{o}$ and the fitted value of $\overline{\tau}$ in each lobe that provides the fit to the``double-humped" SSA region between 250 MHz and 800 MHz (in the quasar rest frame). Another value of $B$ is chosen and the process repeated in order to generate two new values of $N_{\Gamma}$ and $R$ that reproduce the spectral fit. The process is repeated until the solution space of $B$, $N_{\Gamma}$ and $R$ is spanned for each lobe.

\section{Fitting the Data with Leptonic Plasmoid Models}
There are two possible plasma sources for the radio emission in the lobes of 3C82 based on previous applications of these types of plasmoid models to the ejection of relativistic plasma from compact astrophysical objects. Firstly, a turbulent magnetized plasmoid made of electrons and positrons. This was the preferred solution for the major flares in GRS~1915+105 \citep{fen99,pun12}. This will be referred to as a leptonic lobe in the following. Alternatively, a lobe might be a turbulent magnetized plasmoid made of protons and electrons. This was a possible solution for the ejection from the neutron star merger and gravity wave source GW170817 \citep{pun19}. This will be referred to as a protonic lobe in the following. We note that the possibility of protonic lobes in extragalactic radio sources was recently studied in detail \citep{cro18}. It was argued that leptonic lobes are favored in the more luminous FRII radio sources. Leptonic physical models which fit the data in Figure 4 are described in this section. There are two sources of degeneracy in the solutions, the spectral fit itself and the physical model that creates the fitted spectrum. First, the uncertainties in the radio data, the error bars in Figure 4, allow for slightly different spectral functions to fit the data. The physical degeneracies arise from uncertainties in $E_{min}$ and the infinite one dimensional set of solutions described in the last section. Thus, we need some additional insight to guide us toward a plausible physical solution. We consider three physical constraints on the models in order to reduce the degeneracy.
\begin{itemize}
\item Constraint 1: Over long periods of time, we assume that there is an approximate bilateral symmetry in terms of the energy ejected into each jet arm. Thus, we require $E(\mathrm{lm})$ to be approximately equal in both lobes in spite of the fact that the spectral index is not.
\item Constraint 2: The radius of the western lobe should be larger than the eastern lobe based on the X-band image in Figure 3. This is the best data available for making this assessment. The other resolved image, the 5 GHz image from MERLIN agrees, but we noted that $\sim 15\%-20\%$ of the flux was not detected in the eastern lobe due to patchy u-v coverage.
\item Constraint 3: We focus the discussion towards minimum energy configurations, but do not ignore the possibility of deviations from this configuration.
\end{itemize}
\par The degeneracy in the spectral shape will be assessed by the minimization of the residuals and the implications of the supporting physical models noted in the three constraints, above. We address the residuals by considering the excess variance of the fit to the 9 data points that cover the apparent power law from $\nu=1.67$ GHz to $\nu=32.7$ GHz in the quasar rest frame, $\sigma_{\rm{rms}}^{2}$, \citep{nan97},
\begin{equation}
\sigma_{\rm{rms}}^{2}=\frac{1}{N}\sum_{i=1}^{N}
\left[\frac{(S_{i}-f_{i})^2- \sigma_{i}^{2}}{f_{i}^{2}}\right]\; ,
\end{equation}
where, ``i" labels one of the $N$ measured flux densities along the power law, $f_{i}$ is the expected value of this flux density from the fitted curve, $S_{i}$ is the measured flux density and $\sigma_{i}$ is the uncertainty in this measurement. The smaller the value of $\sigma_{\rm{rms}}^{2}$, the better that a particular fit agrees with the data. Note that any fit that is inside the error bars at every data point will produce  $\sigma_{\rm{rms}}^{2}< 0$, i.e. there is less scatter than expected from the estimated uncertainty.

\par As discussed in Section 5.2, our process will first fit the power law (and we record the value of $\sigma_{\rm{rms}}^{2}$ for future comparison with other possible fits). Secondly, we make sure the fit is within the error bars of the four lowest frequency measurements (74 MHz, 151 MHz, 178 MHz and 432 MHz). This determines the SSA opacity in each lobe. One is not guaranteed that there is a solution that fits within the error bars for every pair of lobe power laws and the feedback from fitting the SSA opacity at low frequency typically induces small changes in the power law fits to the individual lobes. Figure 4 shows that this low frequency turnover is too broad to be fit with a single SSA power law. This simply means that the two lobes have different opacity and the two relative maxima broaden the spectral peak. In essence, this is verifying the existence of two different regions of emission at low frequency. We note that typically for SSA power laws and empirically for CSS radio sources that the frequency of the spectral peak of the emitting region scales inversely with the size \citep{van66,mof75,ode98}. Thus, we expect that the $\sim \nu = 300$ MHz peak in the quasar rest frame (near the 74 MHz data point) is emitted from the larger of the two lobes. Indeed, our models consistently find this to be true. Based on Figure 3 and constraint 2, above this favors the western lobe as being responsible for the lowest frequency spectral peak. We consider three different types of fits before choosing our preferred solution.

\subsection{Fit A: Eastern Lobe Maximum $\alpha$, Western Lobe Minimum $\alpha$} We choose the western lobe to be associated with the lowest frequency spectral peak as deduced above. The volume of the eastern lobe is therefore less. We note that as the spectral index of the synchrotron spectrum steepens there are more electrons at lower energy by Equation (6). This creates an increase in energy density that can compensate for the smaller volume in the eastern lobe and help to meet constraint 1, long-term bilateral symmetry in the ejected energy. Similarly, one can choose $\alpha$ as small as possible for the western lobe. This is the strategy for Fit A. In addition, there is a little flexibility in the error bars at $\nu_{o}=151$ MHz and $\nu_{o}=74$ MHz in order to make the SSA opacity as small as possible for the eastern lobe (slightly larger size) and as large as possible for the western lobe (slightly smaller size). Thus, constraint 1, on long-term bilateral symmetry, tends to drive the SSA opacity to the maximum (minimunm) possible value in the western (eastern) lobe that is consistent with the $\nu_{o}=74$ MHz ($\nu_{o}=151$ MHz) lower (upper) error bar in our fits. The result of this strategy is shown in Figure 4 and the upper panel of Figure 5.
\par The relevant details are shown as entry I in Table 3 for direct comparison with other fitting strategies. Table 3 describes the overall fit to the power law region as well as the details of the component fits. Column (1) indicates the name of the particular fit. The details of the combined east lobe plus west lobe solution, the total solution, is described in columns (6) - (10). Column (6) is $\sigma_{\rm{rms}}^{2}$ from Equation (17) for the fit to the power law from $\nu_{o}=432$ MHz to $\nu_{o}=8.44$ GHz in the observer's frame. Since  $\sigma_{\rm{rms}}^{2}$ is driven largely by the outlier $\nu_{o}=686$ MHz GBT data, we remove this point from the $\sigma_{\rm{rms}}^{2}$ calculation in Column (7) in order to test whether this one point is changing our fit to fit comparison. Column (8) tabulates the ratio of the lepto-magnetic energy ($E(\rm{lm})$ from Equation(12)) of the east lobe to $E(\rm{lm})$ of the west lobe assuming a minimum energy solution in both lobes as the source of the resultant spectrum. The next column is the total $E(\rm{lm})$ and the last column identifies the largest lobe based on the physical model. Indented below these data are the details of the east and west lobe fits, the peak frequency, the luminosity at the spectral peak and the power law spectral index. The only entries for the individual lobes are columns (3)-(5), the other columns are not applicable, N/A.
\par Figure 6 describes the physical parameters of the model. The top panels of Figure 6 show the dependence of $E(\mathrm{lm})$ in Equation (12) on $R$ (the lobe radius) for the leptonic lobes that produce the spectra in Figure 4. The middle row shows the dependence of $E(\mathrm{lm})$ on the total number of particles in the lobe, $\mathcal{N}_{e}$. The bottom row shows the dependence on the turbulent magnetic field strength.

\subsection{Fit B: Eastern Lobe Minimum $\alpha$, Western Lobe Maximum $\alpha$} Alternatively, Fit B adjusts $\alpha$ to be as flat as possible in the eastern lobe and $\alpha$ to be as steep as possible in the western lobe. The fit is shown in the top frame of Figure 5. The corresponding minimum energy model is entry II in Table 3. Note that this choice produces less total flux in the $\nu_{o}=432$ MHz to $\nu_{o}=686$ MHz range compared to Fit A in Figure 4. This results in an increase of $\sigma_{\rm{rms}}^{2}$ of the fit relative to Fit A in Columns (6) and (7). The choice also increases $E(\rm{lm})$ of the western lobe and decreases $E(\rm{lm})$ in the eastern lobe for the minimum energy solution. Column (8) shows that the spectral index difference between the lobes is no longer large enough to compensate for the volume difference in the lobes and the solution deviates significantly from bilateral symmetry in the ejected energy. Thus, we consider this solution less plausible than Fit A, but it is not formally excluded by any of the data.
\subsection{Fit C: Reversed Lobe Assignments}
Finally, we consider the possibility that our lobe assignments with the SSA peaks is backwards in Fits A and B. Fit C reverses these assignments and is shown in the bottom frame of Figure 5. The corresponding minimum energy model is entry III in Table 3. This fit has the lowest flux in the $\nu_{o}=432$ MHz to $\nu_{o}=686$ MHz range and the largest $\sigma_{\rm{rms}}^{2}$. It is also the farthest from bilateral energy ejection symmetry in Table 3. Furthermore, the fit is below the lower bound VLBA data at $\nu_{o}=327$ MHz. For these reasons, it is the least plausible fit.
\subsection{A non-Minimum Energy Model for Fit A}
In this section, we consider a different model of the lobes in order to understand the effects of abandoning the minimum energy and the $E_{min}\approx m_{e}c^{2}$ assumptions. This solution radiates the same spectrum as Fit A in Figure 4. The eastern lobe solution is the same as the minimum energy solution described in Figure 6. The western lobe solution is altered from that in Figure 6 and the previous subsections. We now choose $E_{min}\approx 5 m_{e}c^{2}$. The solutions are still governed by constraints (1) and (2) listed at the beginning of Section 6. However, the western lobe violates minimum energy, constraint (3).
\par The characteristics of the solution are plotted in Figure 7. Note that the minimum energy in the two lobes is drastically different in Figure 7. There is no minimum energy solution that fulfills constraint (1). We choose a solution with $U_{B}>U_{e}$ in the western lobe. The horizontal dashed line in the top panel of Figure 7 connects the minimum energy solution in the eastern lobe to the magnetically dominated condition in the western lobe, under the time-averaged, bilateral symmetry condition that is embodied in constraint (1): $E(\mathrm{lm})(\rm{western\, lobe})\approx E(\mathrm{lm})(\rm{eastern\, lobe})$. The blue dot designates the location of this solution in the three frames of Figure 7. The condition that $\approx 90\%$ of the lobe energy is in the magnetic component might seem extreme and not plausible. However, a detailed analysis of the ejection of large leptonic plasmoids (using the same modeling as here) in the Galactic black hole, GRS~1915+105 showed that the plasmoids evolve from being magnetically dominated towards equipartition as they propagate. CSS sources are young for an FRII radio source, so it is not unreasonable that it has not relaxed to a minimum energy configuration. The fact that the eastern lobe has reached a minimum energy configuration and the western lobe has not yet reached a minimum energy configuration might be a consequence of a different life history during their propagation from the central engine. The details of the solution are described as entry IV in Table 3.
\par This solution suggests studying a model in which $E_{min}\approx 5 m_{e}c^{2}$ in both lobes. If we assume minimum energy in both lobes, this model produces a minimum energy 2.5 times larger in the west lobe than the east lobe. This is a consequence of the east lobe spectrum being much steeper, the low energy cutoff removes more leptons from components with steeper spectral indices. It is clear that due to the different values of $\alpha$, assigning the same $E_{min}\neq m_{e}c^{2}$ in each lobe, does not produce solutions with the same minimum energy. Only when both lobes have $E_{min}\approx m_{e}c^{2}$ can equal minimum energy be achieved in our models for 3C 82.

\subsection{Effects of Lobe Doppler Factor in the Minimum Energy Model for Fit A} We consider the model of Section 6.1 for Fit A with the modification that we do not assume $\delta =1$. Both lobes are at minimum energy with $E_{min}\approx m_{e}c^{2}$. Based on the discussion in Section 5, we consider a single lobe speed that is a crude attempt to average the hot spot speed, $0.05c < v_{hs}<0.1c $ and the diffuse lobe plasma velocity, $v_{diffuse} \ll v_{hs}$ (neither of which is observed or well known). We choose the entire western (eastern) lobe to advance (recede) at a speed 0.05c. Assuming that the western lobe is the advancing lobe is consistent with it being brighter and being adjacent to the putative jet. $E_{lm}$ of the eastern lobe increases by $\sim 10\%$ from $6.76 \times 10^{59}$ ergs to $7.62 \times 10^{59}$ ergs. By contrast $E_{lm}$ of the western lobe decreases $\sim 10\%$ from $8.70 \times 10^{59}$ ergs to $7.84 \times 10^{59}$ ergs. The size of the approaching (receding) lobe gets slightly smaller (larger), most of the energy change is accomplished by a density change. The details of the model are found in entry V in Table 3. Notice that the system is much closer to bilateral symmetry in the total energy ejected into each jet arm than entry I of Table 3. This would appear to be an improvement to the accuracy of the model. We also added the case where the lobes advance at 0.1c as entry VI in Table 3. Based on entries V and VI in Table 3, the Doppler enhancement has no effect on the total energy in the system for these modest velocities.

\subsection{Comparison and Contrast of the Fits and Models}
\par Fit A has the lowest $\sigma_{\rm{rms}}^{2}$ for the fit to the power law, columns (6) and (7) of Table 3. It is the best fit of any two component SSA model. It is the closest to having bilateral symmetry in the ejected energy (constraint 1, above), column (8) and it has the west lobe larger than the east lobe (constraint 2, above) per column (10). For these reasons we consider this the most plausible choice for the best fit within the context of the minimum energy assumption. We note why $\sigma_{\rm{rms}}^{2}$ for the fit to the power law is lower in this case. The $\nu_{o}=432$ MHz, $\nu_{o} =602 $ MHz and the $\nu_{o} =686 $ MHz flux densities require a significant contribution from the smaller lobe in order to be attained by the sum of the two lobe flux densities. This is best achieved by moving the spectral peak to lower frequency and increasing the flux at low frequency by making the spectral index as steep as possible.
\par {Table 3 indicates that the process of adjusting the individual lobe spectral indices so as to minimize the excess variance of the power law fit to the radio data, naturally drives the solution to one in which there is bilateral symmetry in the ejected energy in the minimum energy limit. The excess variance is very sensitive to the fitting of the high frequency edge of the SSA local maximum. This is the most complicated and constraining region of the fit, both lobes contribute significantly and the fitting of this spectral hump is constrained by the nearby $\nu_{o} =432$ MHz, $\nu_{o} =602$ MHz and the $\nu_{o} =686$ data. This is the region that produces most of the excess variance in Fits B and C. But, Table 3 verifies that the identification of Fit A as the best fit is robust, it still holds even if the apparent outlier data at $\nu_{o} =686$ MHz is removed.
\par The solutions depend critically on the $\nu_{o}=151$ MHz and $\nu_{o}=74$ MHz data that define the SSA region. We corroborated the $\nu_{o}=151$ MHz 6C data with the GMRT $\nu_{o}=150$ MHz data as discussed above. We also downloaded the VLA B-array $\nu_{o}=74$ MHz image\footnote{https://www.cv.nrao.edu/vlss/VLSSlist.shtml} and there are no confusing sources in the field of view, nor is there any evidence of strong sidelobes resulting from u-v coverage issues \citep{lan12}. We can find no reason to doubt the accuracy of these data.

\begin{figure*}
\begin{center}
\includegraphics[width= 0.6\textwidth,angle =0]{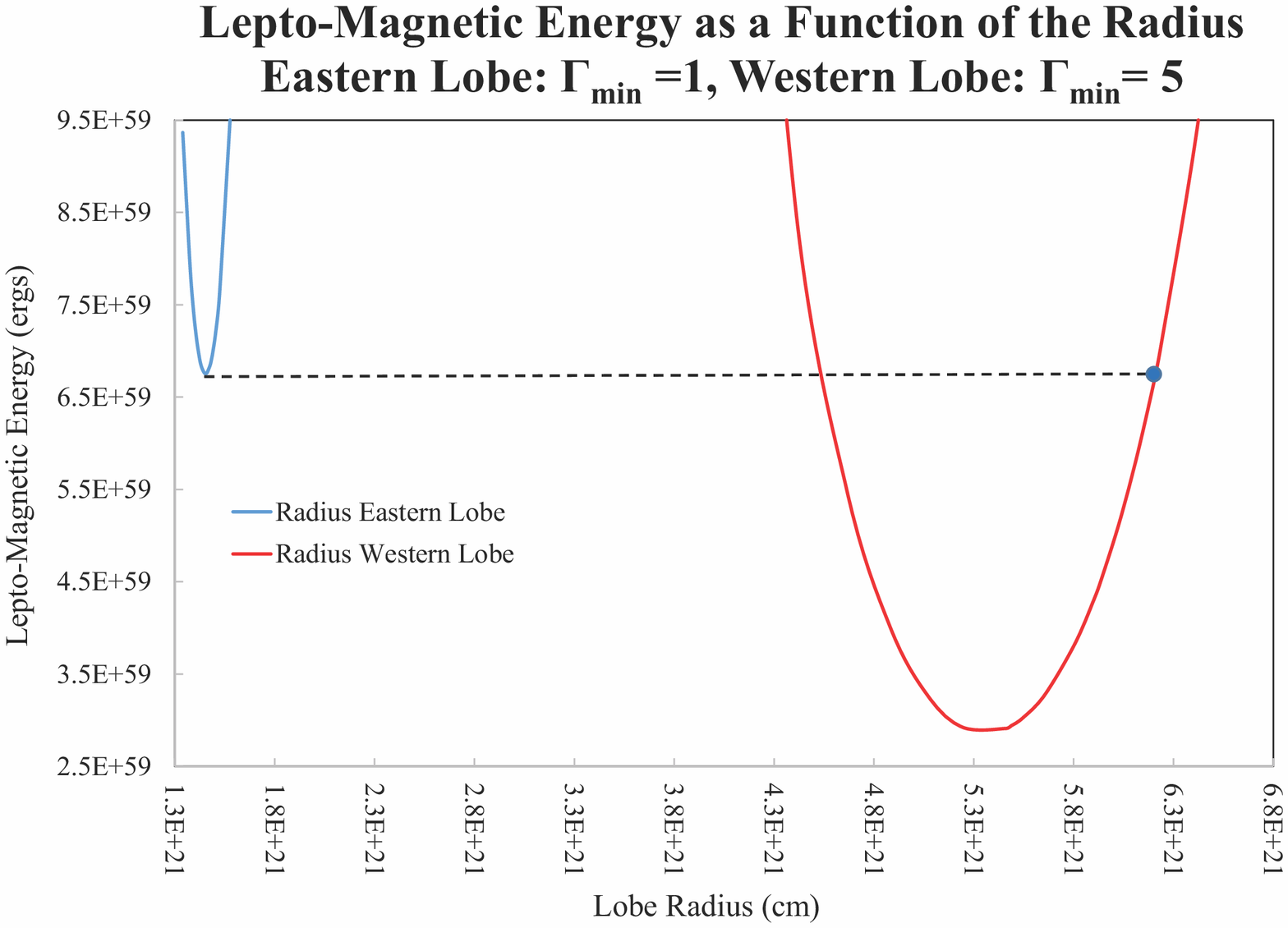}
\includegraphics[width= 0.45\textwidth,angle =0]{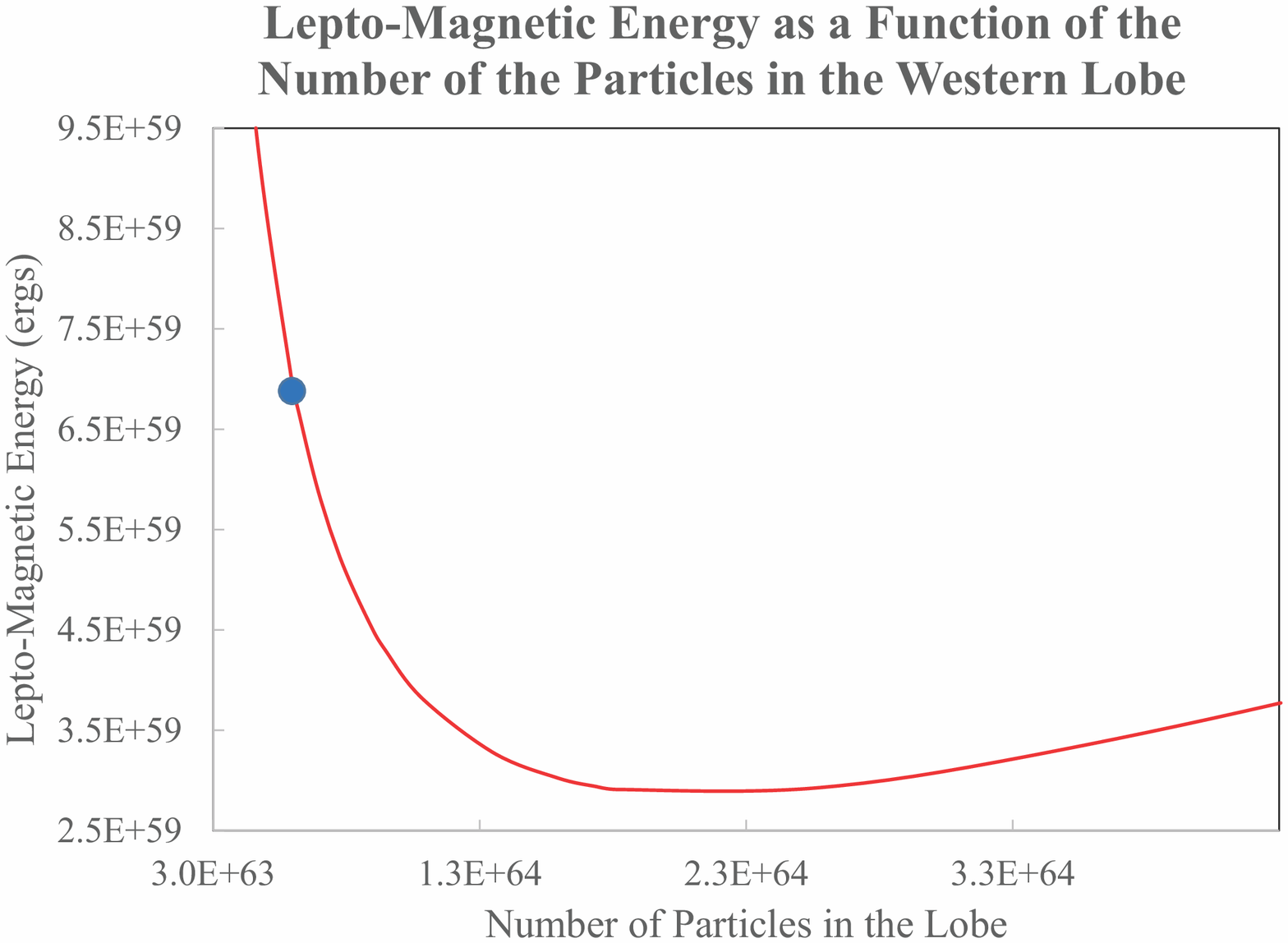}
\includegraphics[width= 0.45\textwidth,angle =0]{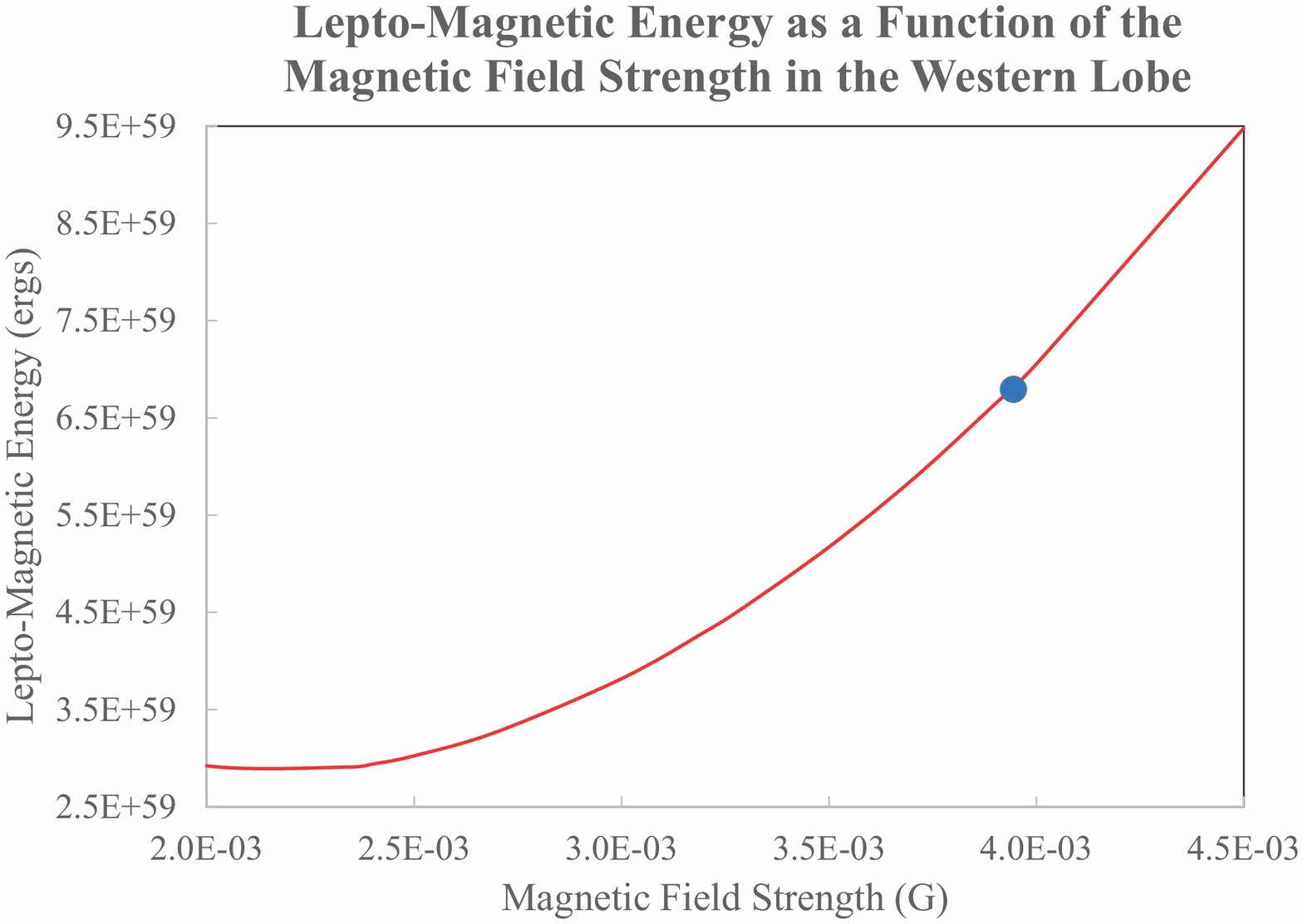}
\caption{The same data that is in Figure 6 that is adapted to the non-minimum energy solution of Section 6.4. The eastern lobe solution is the same as in Figure 6 (a minimum energy solution) and the western lobe is magnetically dominated. The blue dot shows the location of the magnetically dominated solution on each plot.}
\end{center}
\end{figure*}

\subsection{The Uniqueness of the $E_{min}\approx m_{e}c^{2}$ Assumption}
Considering the six solutions in Table 3, it is clear that there is only one solution with all three of the following properties:
\begin{enumerate}
\item Both lobes are near the minimum energy state (constraint 3),
\item The magneto-leptonic energy, $E_{lm}$, is approximately the same in both lobes (constraint 1):
 $E_{lm}(\rm{West\, Lobe}) \approx E_{lm}(\rm{East\, Lobe})$,
\item The low energy cutoff to the lepton power law is the same in both lobes:
 $E_{min}(\rm{West\, Lobe}) = E_{min}(\rm{East\, Lobe})$.
\end{enumerate}
That solution has $E_{min}(\rm{West\, Lobe}) = E_{min}(\rm{East\, Lobe}) = m_{e}c^{2}$ with Fit A. This solution is unique and has the properties that we posited as relevant at the start of this section. It also has the esthetic that there are no unexplained spectral breaks or low energy cutoffs in the lepton spectra. Of the models in Table 3 that have these properties, the model with a lobe advance speeds of 0.05c (entry V from Table 3) in Section 6.5 might be preferred since it has almost exact bilateral symmetry in the energy output.

\section{Converting Stored Lobe Energy into Jet Power} There is a direct physical connection between $E(\mathrm{lm})$ and the long-term time-averaged power delivered to the radio lobes, $\overline{Q}$. In this section, we crudely estimate this relation for 3C 82. If the time for the lobes to expand to their current separation is $T$, in the frame of reference of the quasar, then the intrinsic jet power is approximately
\begin{equation}
\overline{Q} \approx E(\mathrm{lm})/T
\end{equation}

\subsection{Estimating the Expansion Time, $T$}The main goal of this section will be to find an estimate of the mean lobe advance speed, $v_{\rm{adv}}$. that is applicable to 3C 82. There is no evidence of motion of the lobes in the images, so one must use indirect means. In particular, we will use the statistics of estimates of $v_{\rm{adv}}$ for ``similar" objects. The main method that is implemented to this end is jet arm length asymmetry. This method requires identifying the approaching jet and knowing the core position. Even so, the method cannot be reliably applied to single objects since the results are heavily skewed by intrinsic asymmetry in the ambient environment. If one does have perfect bilateral symmetry in emission and in the enveloping medium one can formulate the arm length ratio of approaching lobe, $L_{\rm{app}}$, to receding lobe, $L_{\rm{rec}}$, as \citep{gin69,sch95}
\begin{eqnarray}
&& R_{asym} = \frac{L_{\rm{app}}}{L_{\rm{rec}}} = \frac{1+(v_{\rm{adv}}/c) \cos{\theta}}{1-(v_{\rm{adv}}/c) \cos{\theta}} \;, \rm{where}\nonumber \\
&& L_{\rm{app}}=\frac{(v_{\rm{adv}}T)\sin{\theta}}{1-(v_{\rm{adv}}/c) \cos{\theta}} \;: \; L_{\rm{rec}}=\frac{(v_{\rm{adv}}T) \sin{\theta}}{1+(v_{\rm{adv}}/c)\cos{\theta}}\;,
\end{eqnarray}
where $T$ is the time measured in the quasar rest frame and projected length on the sky plane of an earth observer (corrected for cosmological effects) is $\mathcal{L}\equiv L_{\rm{app}}+L_{\rm{rec}}$.
Since this formula is not reliably applicable to single sources, the preferred method is to define samples of ``similar" objects and look at the statistical distribution of $v_{\rm{adv}}$ \citep{sch95}. But, how does one define the notion ``similar" in the case of 3C82? The jet propagation has been studied in terms of simple self-similar models which assume an ambient density that scales like $n_{\rm{ambient}}=n_{o}r^{-\psi}$, where $r$ is the distance from the quasar \citep{kai97,wil99}. The following relevant scalings have been shown in Equations (10) and (11) of \citet{wil99}, respectively
\begin{eqnarray}
&& v_{\rm{adv}} \propto T^{(\psi-2)/(5-\psi)} \overline{Q}^{1/(5-\psi)}\\
&& v_{\rm{adv}} \propto D^{(\psi-2)/3} \overline{Q}^{[\psi-1]/[3(5-\psi)]}\;,
\end{eqnarray}
where $D$ is the size of the source projected on the sky plane. One problem is that we do not know $\psi$, but this has been estimated as 1.5 \citep{wil99}. This value is consistent with azimuthally averaged density profiles estimated for elliptical galaxies \citep{mat03}. However, for a jet traversing a given direction through the galaxy, a simple power law is a crude approximation. In any event, Equations (17) and (18), indicate that for these simple models, $v_{\rm{adv}}$ increases with jet power and decreases (decelerates) with distance from the quasar. 3C 82 is at the high end of the $\overline{Q}$ distribution and is smaller than most of the FRII quasars in the 3CRR catalog \citep{lai83}. In terms of the first requirement, the most luminous quasar sample considered in \citet{sch95} was the 3CRR sample, for which his Monte Carlo simulations found a median $v_{\rm{adv}}=0.115 c$. However, these objects have a median length of $\gtrsim 110$ kpc versus the 11 kpc found for 3C 82. For $\psi=1.5$ in Equation (17), $v_{\rm{adv}} \sim D^{-0.167}\overline{Q}^{0.047}$. We will find that $\overline{Q}$ of 3C 82 is $\sim 100$ times that of the median 3CRR source. Using this relation and the 3CRR analysis of \citet{sch95}, we estimate a most likely value for 3C82 based on its size and luminosity of
\begin{equation}
v_{\rm{adv}} \approx 0.115c (110/11)^{0.167}(100)^{0.047} \approx 0.21 c
\end{equation}
\par In order to corroborate the estimate in Equation (19), we assembled a sample of CSS quasars of similar size ($2\rm{kpc}< D< 25\rm{kpc}$). We require very straight jets that would be indicative of negligible interaction with the enveloping media. We need a tight mathematical constraint on straightness. To this end, we require: if the vertex of a cone with a half angle $15^{\circ}$ is placed on one of the lobes, then the other lobe, the jet and the core in all high dynamic range radio images can be contained within the cone. We also require a very high lobe luminosity, thus we restricted the sample to 3C objects. We eliminated objects previously identified as CSS that turned out to be much larger based on higher dynamic range imaging and objects that turned out to be powerful blazars with strong lobe emission that only appeared small due to the polar line of sight. We then computed $R_{asym}$ from Equation (16) using the highest sensitivity and dynamic range images in the public domain. We estimated the arm lengths based on the methods of \citet{sch95}. The results are tabulated in Table 4.
\begin{table}
\caption{Arm Length Asymmetry in Straight 3C CSS Quasars}
{\footnotesize\begin{tabular}{ccccc} \tableline\rule{0mm}{3mm}
Source  & Linear Size (kpc) & $R_{asym}$  & $v_{\rm{adv}}$/c\tablenotemark{a} & References\\
\tableline \rule{0mm}{3mm}
3C 138  & 4.9 & 1.66 & $0.30\pm 0.03$ & \citep{aku95} \\
3C 186  & 18.1 & 1.21 & $0.11^{+0.02}_{-0.01}$ & \citep{aku95}\\
3C 277.1  & 17.2 & 1.83 & $0.34^{+0.04}_{-0.03}$ & \citep{rei95,pea85}\\
3C 298  & 9.4 & 1.87 & $0.35^{+0.04}_{-0.03}$ & \citep{aku95}
\end{tabular}}
\tablenotetext{a}{Assumes $ 20^{\circ} < \rm{LOS}< 40^{\circ}$.}
\end{table}
The advantage of restricting the CSS sources to quasars is that it eliminates the very oblique LOS of radio galaxies. The LOS to the jet in quasars is believed to be $<45^{\circ}$ with an average of $30^{\circ}$ \citep{bar89}. By eliminating blazar LOSs, $<10^{\circ}$, we have a narrow range of LOS that will provide only a few percent variation of the $v_{\rm{adv}}$ in column (4). The values in Table 4 are slightly larger than what was expected from Equation (19). Based on Table 4, the 3CRR analysis of \citet{sch95} and Equation (19), we conclude that
\begin{equation}
v_{\rm{adv}} \approx 0.2 c \pm 0.1 c\;,
\end{equation}
should cover the lobe advance speeds that might be applicable to 3C 82. Assuming, bilateral symmetry and a LOS $\approx 30^{\circ}$, Equations (16) and (20) yield a loose bound on $T$,
\begin{eqnarray}
&& T = \frac{\mathcal{L}\left[1 - (\frac{v_{\rm{adv}\cos{\theta}}}{c})^{2}\right]} {2v_{\rm{adv}}\sin{\theta}}\approx 11.3 \, \rm{kpc} /v_{\rm{adv}}, \,\rm{or}\,\; \nonumber\\
&&  3.86 \times 10^{12} \rm{sec} < T< 1.16\times 10^{13} \rm{sec} \;.
\end{eqnarray}
The uncertainty in $T$ in Equation (21) is driven by the large uncertainty in $v_{\rm{adv}}$ in Equation (20) that is estimated from the spread in values in Table 4.

\subsection{Estimates of the Jet Power}
From the minimum energy solution corresponding to Fit A in Table 3 and Figure 6, the total lepto-magnetic energy stored in both lobes is approximately
\begin{equation}
E(\mathrm{lm})\approx 1.55 \times 10^{60} \rm{ergs}\;.
\end{equation}
Equations (15) and (21) combined with Equation (22) implies a very large lower bound on the jet power of
\begin{equation}
\overline{Q}> 1.33 \times 10^{47} \rm{ergs/sec}\;.
\end{equation}
This is a conservative lower bound because it does not include work done by the expansion into the ambient medium which can be of comparable magnitude \citep{wil99}. There is no observational data that can reliably constrain this for 3C82.
Ignoring the contribution from work on the external environment, Equations (15) and (21) imply
\begin{equation}
\overline{Q}\approx 2.66\times 10^{47} \rm{ergs/sec} \pm 1.33 \times 10^{47} \rm{ergs/sec}\;.
\end{equation}
\textbf This value is similar to that obtained by the same methods applied to the non-minimum energy solution of Section 6.4, $\overline{Q}\approx 2.32\times 10^{47} \rm{ergs/sec} \pm 1.16 \times 10^{47} \rm{ergs/sec}$. Similarly, from Table 3 for the minimum energy solution for Fit B on finds $\overline{Q}\approx 2.32\times 10^{47} \rm{ergs/sec} \pm 1.16 \times 10^{47} \rm{ergs/sec}$. Thus, for a wide range of assumptions the estimated jet power is very similar.
\par It is of interest to compare this with more traditional estimates of $\overline{Q}$. The spectral luminosity at 151 MHz per steradian, $L_{151}$, provides a surrogate for the luminosity of the radio lobes in a method that assumes a relaxed classical double radio source \citep{wil99}. The method assumes a low frequency cut off at 10
MHz, the jet axis is $60^{\circ}$ to the line of sight, there is no
protonic contribution and 100\% filling factor. The plasma is near minimum energy and a quantity, $f$,
is introduced to account for deviations of actual radio lobes from these assumptions
as well as energy lost expanding the lobe into the external medium,
back flow from the head of the lobe and kinetic turbulence. $\overline{Q}$ as
a function of $f$ and $L_{151}$ is plotted in Figure 7 of \citet{wil99},
\begin{equation}
\overline{Q} \approx 3.8\times10^{45} f L_{151}^{6/7} \rm{ergs/s}\;,
\end{equation}
Note that exponent on $f$ is 1 not 3/2 as was previously reported \citep{pun18}. The quantity
$f$ was estimated to be in the range of $10<f<20$ for most FRII radio sources
\citep{blu00}. Using $L_{151}\approx 5.8 \times 10^{28} W\rm{Hz}^{-1}\rm{sr}^{-1}$ from Figure 4, Equation (31) and $10<f<20$
\begin{equation}
\overline{Q} = 2.07\pm 0.69\times 10^{47} \rm{erg/s} \;.
\end{equation}
Note the close agreement between our estimate in Equation (24) and Equation (26).

\begin{table}
    \caption{Quasars with time-averaged Jet Powers Exceeding $10^{47}$ ergs/sec }
    \tiny
    \centering
    \begin{tabular}{ccccccccc}
        \hline
        (1) & (2)  & (3)  & (4) & (5)  & (6) & (7)  &(8) & (9) \\
        Quasar  & z  & $L_{151}$  &  $\overline{Q}$ &Method & Reference & $L_{\rm{bol}}$ & Reference  & $\overline{Q}/L_{\rm{bol}}$     \\
        &          &  $10^{28} W\rm{Hz}^{-1}\rm{sr}^{-1}$        &  $10^{47}$ ergs/sec &          &      &  $10^{46}$ ergs/sec    &      Spectrum       &              \\
        \hline
        3C298 & 1.44  & 6.04\tablenotemark{a} & $2.10 \pm 0.70$ \tablenotemark{b} & Eqn. 31 & This Paper & 7.20 & HST FOS G270H  &$2.92 \pm 0.97$ \\

        3C82       & 2.87  & 5.94  & $2.07 \pm 0.69$ &  Eqn. 31 & This Paper & 4.88 & This Paper  &$4.24\pm1.41 $ \\
        3C82       & 2.87  & 5.94  & $2.66 \pm 1.33 $ & SSA Plasmoid & Sections 4-7 & 4.88\tablenotemark{c} & This Paper  &$5.45\pm 2.73 $\tablenotemark{c} \\

        3C9       & 2.01  & 5.01  & $1.79 \pm 0.60$  &  Eqn. 31 & This Paper & 8.97 & SDSS    &$2.00\pm0.67 $\\\

        4C+11.45       & 2.18  & 3.08   & $1.16 \pm 0.39$  &  Eqn. 31 & This Paper & 2.11\tablenotemark{d} & SDSS  & $5.41 \pm 1.80 $  \\

        PKS 0438-436       & 2.86  & 3.00\tablenotemark{d}   & $1.15 \pm 0.39$\tablenotemark{e} &  Eqn. 31 & \citet{pun18} & 4.7 & \citet{pun18}    & $2.45\pm0.82$

    \end{tabular}
\footnotesize{\tablenotetext{a}{The lobe luminosity is estimated from higher resolution, higher frequency images due to the existence of a very prominent radio core and jet.}\tablenotetext{b}{The use of Equation (25) for 3C298 is uncertain due to the small size and the luminous core plus jet. $\overline{Q}$ needs to be verified by independent means, as we did for 3C 82 in this paper.}
\tablenotetext{c}{This estimate does not include all the sources of uncertainty in $L_{\rm{bol}}$ that were developed in this paper (see Section 7.2 for such an estimate) in order to provide a consistent comparison to the other quasars.}
\tablenotetext{d}{Based on two SDSS spectra. The continuum is 0.45 of the level measured in \citet{bar90}, which they said was ``roughly calibrated".}
\tablenotetext{e}{$L_{151}$ and $\overline{Q}$ are derived from the un-beamed radio lobes. The Doppler boosted jets and blazar core are not included to improve the accuracy of Equation (25) \citet{pun18}}}
\end{table}
\clearpage
\par In order to place this jet power estimate in context, we compare this result with other quasars with extremely large values of $L_{151}$ in Table 5. The first column is the name of the quasar followed by the redshift. Column (3) is $L_{151}$ of the radio lobes, this is the appropriate luminosity to use in Equation (25) in order to calculate $\overline{Q}$ \citep{pun18}. Columns (4) and (5) are $\overline{Q}$ and the method used for estimation. Column (6) is the reference to where this was done. Columns (7) and (8) are an estimate of $L_{\rm{bol}}$ from Equations (1) and (2) with a reference to the spectrum that was used. The last column is an estimate of $\overline{Q}/L_{\rm{bol}}$ from columns (4) and (7). Table 5 shows that 3C 82 has the most luminous radio lobes, within observational uncertainty, and one of the largest values of jet dominance, $\overline{Q}/L_{\rm{bol}}=5.45\pm 2.73$. This expression used the average $L_{\rm{bol}}$ from Section 2. If we include the variability over the three epochs and uncertainty in Equation (1), $3.2 \times 10^{46} \rm{ergs/sec}<L_{\rm{bol}}< 5.8 \times 10^{46} \rm{ergs/sec}$, with the uncertainty in $\overline{Q}$ from Equation (24) then we have a more rigorous estimate, $\overline{Q}/L_{\rm{bol}}=5.91\pm 3.41$, where the uncertainties are added in quadrature. The source, 3C 298, is likely less powerful than indicated in the table since it has a strong radio core and strong knots in the jet indicating increased dissipation relative to that of relaxed radio lobes \citep{lud98}. Thus, we could be over-estimating the jet power based on Equation (25) \citep{wil99}. By contrast, there are no detected strong knots or radio core in 3C 82, just the very steep spectrum lobes with no highly delineated hot spots. It is conspicuous that all the powerful radio sources in Table 5 (the most powerful known radio quasars) have $\overline{Q}/L_{\rm{bol}}\gtrsim 1$.
\subsection{Protonic Lobes} In principle, the positive charges in the ionized lobes can be protonic matter instead of positronic matter. Based on Equation (11) and Equation (19), the kinetic energy of the lobes would be much larger than $E(\mathrm{lm})$. The spectrum in Figure 4 is the same and is created by electrons in the magnetic field. Using these same methods to describe a relativistic plasmoid ejection it was determined that this was a viable solution for GW170817, and associated gamma ray burst, GRB 170817A \citep{pun19}. However, in this case consider the mass stored in the lobes based on Figure 6 for the minimum energy, minimum $E(\mathrm{lm})$, solution, $M_{\rm{lobe}} = 5.44\times 10^{8} M_{\odot}$. We can compare this to the accreted mass,
$M_{\rm{acc}}= T \dot{M}= T L_{\rm{bol}}/(\eta c^{2})$, where $\eta$ is the radiative efficiency of the accretion flow which we take to be 10\% and $\dot{M}$ is the accretion rate \citep{nov73}. From $L_{\rm{bol}}$ in Equation (1) and the average value of the jet age in Equation (21), the accreted mass during the jet lifetime is
$M_{\rm{acc}}= 1.58\times 10^{6} M_{\odot}$. Thus, in the minimum energy solution $M_{\rm{lobe}}/M_{\rm{acc}} =341$. For the non-minimum energy solution in Section 6.4, $M_{\rm{lobe}}/M_{\rm{acc}} = 165 $. These solutions require two orders of magnitude more mass to be transported to the radio lobes than is accreted toward the supermassive central black hole. For this reason, the protonic solutions for the lobes in 3C 82 are disfavored.

\section{The High Ionization Wind} Table 1 shows that the CIV BEL has both a high redshift (red VBC) component and a high blue shift component (BLUE). However, the BLUE has 2.3 times the luminosity of the red component. This balance is very unusual for Population B quasars and radio loud quasars, yet it is still seen occasionally \citep{sul15}. The strong blue-shifted component is believed to arise from an out-flowing wind \citep{bro94,bro96,mur95}. The wind in the case of 3C 82 must be very high ionization compared to other radio loud quasars with a dominant BLUE component since there is no detected BLUE in the lower ionization line, CIII] \citep{sul15}. In this section, we estimate the power in such a high ionization wind using two models, the continuous wind model of \citet{mur95} and an outflow in a system of clouds in pressure equilibrium, under the combined effect of radiation and gravity \citep{net10,mar17}.

\subsection{Line Driven Wind}The line driven continuous wind has the advantage that it does not require a cloud confinement mechanism. It avoids large optical depths since the wind is accelerating, gas is constantly being Doppler blue-shifted relative to the accretion disk and the inner gas. The acceleration scale length is
$\sim 2 \times 10^{13}$ cm, basically the ad hoc cloud dimension in other models \citep{mur95}. The outflowing wind is more highly ionized than cloud models. This is quantified in terms of the ionization parameter, $U$,
\begin{equation}
U = \frac{\int_{3.3\times 10^{15}\rm{Hz}}^{\infty}(L_{\nu}/\rm{h}\nu) \, d\nu}{4\pi r^{2}n_{w}\rm{c}}\;,
\end{equation}
where $n_{w}$ is the hydrogen number density in the wind. In the line emitting region $U \approx 1-10$. Most of the emission comes from the base of the wind where $U$ is smaller, $U$ increases as the wind accelerates away from the source. Based on Figure 1, we approximate the flux density
$F_{\lambda}(\lambda = 916 \rm{\AA}) \approx 2 \times 10^{-15}$ ergs/sec/$\rm{cm}^{2}-\AA$ in the quasar rest frame. The spectral index in the extreme ultraviolet in frequency space is chosen to be very steep, $\alpha_{\nu} = 2$, as is indicated for powerful radio loud quasars \citep{zhe97,tel02,pun15}. Thus, the numerator in Equation (27), the flux of ionizing photons, $N_{\rm{ion}}=2 \times 10^{56} \rm{sec}^{-1}$. From Equation (27), we get the following constraint on the CIV emitting gas
\begin{equation}
4\pi r^{2}n_{w} \approx 2.2 \times 10^{45} \rm{cm}^{-1}\frac{3}{U}\;.
\end{equation}
This equation is useful because the wind power (kinetic energy flux) is
\begin{equation}
P_{w} =4\pi r^{2}n_{w}f_{c} m_{p} \frac{1}{2}v_{w}^{3}\;,
\end{equation}
where $m_{p}$ is the mass of a proton, $v_{w}=2618$ km/sec is from the peak of the CIV BLUE Gaussian component in Table 1 and the wind covering factor is $f_{c} \approx 0.04$ \citep{mur95}. This is based on the estimate that quasars are viewed with a LOS within $45^{\circ}$ of the accretion disk normal \citep{bar89}. If broad absorption line quasars are those viewed through the wind and their rate of occurrence is $\approx 10\%$ then $f_{c} \approx 0.04$. Thus, Equations (28) and (29) yield
\begin{equation}
P_{w} =1.36\times 10^{45} \frac{3}{U}\frac{f_{c}}{0.04} \rm{ergs/sec}\;.
\end{equation}
\subsection{Cloud Outflow}In the cloud outflow model, the force that drives the clouds outward is radiation pressure. Thus, high luminosity accretion flows produce the most luminous BLUE outflow components. The wind power is given by \citep{mar17}:
\begin{eqnarray}
&& P_{w} =2.44\times 10^{44}k^{2}\frac{L(\rm{BLUE})}{10^{45}\rm{ergs/sec}}\left[\frac{v_{\rm{clouds}}}{5000\, \rm{km/sec}}\right]^{3}\frac{1 \rm{pc}}{r_{BLR}}\frac{10^{9}\, \rm{cm}^{-3}}{n_{\rm{clouds}}}\frac{5Z_{\odot}}{Z}\rm{ergs/sec}\;, \\
&& k \equiv \frac{v_{\rm{terminal}}}{v_{\rm{clouds}}}
\end{eqnarray}
where $v_{\rm{terminal}}$ is the terminal velocity of the outflow, $v_{\rm{clouds}} = 2618$ km/sec is the BLUE cloud velocity from Table 1,
$L(\rm{BLUE})=1.78 \times {44}$ ergs/sec is the luminosity of the BLUE from Table 1, $r_{BLR}$ is the distance to the BLUE and the metallicity of the clouds is $Z$. Reverberation mapping of high redshift quasars of similar luminosity indicates \citep{lir18}
\begin{equation}
\frac{r_{BLR}}{10 \rm{light-days}} = (0.22\pm 0.10)\left[\frac{\lambda L_{\lambda}(1345\,\rm{\AA})}{10^{43}\rm{ergs/sec}}\right]^{0.48\pm0.08}\;,
\end{equation}
However, it is not clear if this can be applied to a wind driven BEL region. Based on CLOUDY models we expect that $U\gtrsim 1$ will provide a strong CIV and a very weak CIII] as desired. From modeling of cloud dynamics using the simulations in \citet{net10}, we find that a hydrogen column density of $N_{H} \lesssim 10^{22} \rm{cm}^{-2}$ is required for efficient acceleration by the radiation field of the quasar. A column density of $N_{H} \approx 10^{23} \rm{cm}^{-2}$ will not be efficiently accelerated. Thus, we have a constraint on the model $N_{H} \lesssim 10^{22} \rm{cm}^{-2}$ that is used in combination with an assumed cloud size of $\sim 10^{13}$ cm, $U\approx 1$ and $N_{\rm{ion}}=2 \times 10^{56} \rm{sec}^{-1}$ as was estimated for 3C82, above. For $N_{H} \approx 2.7 \times 10^{22} \rm{cm}^{-2}$,
$r_{BLR} \approx 4.4 \times 10^{17}$ cm, consistent with the high end of the range of reverberation estimates in Equation (33). For these parameters, the models of outflow in a system of clouds in pressure equilibrium, under the combined effect of radiation and gravity yield $k\approx 10$. From which we can estimate the wind power in Equation (31) for the case of 3C 82
\begin{equation}
P_{w} \approx 1.12\times 10^{45}\left[\frac{k}{10}\right]^{2}\frac{5Z_{\odot}}{Z}\; \rm{ergs/sec} \;.
\end{equation}
\subsection{Wind Energetics} The fundamental nature of the high ionization outflow is not well understood, it could be a continuous line driven wind or an outflow of clouds. The details of both scenario have significant uncertainty as discussed in the previous two subsections. However, both models yield estimates of
$P_{w} \gtrsim 10^{45}$ ergs/sec. The wind kinetic power is approximately ten times the line luminosity. Note that the time averaged jet power is two orders of magnitude larger than the wind power.
\section{Summary and Conclusion}
This study is the first in-depth investigation of the quasar, 3C 82, which is a member of the CSS class of radio sources. We obtained the first high SNR optical (UV rest-frame) spectrum. The first VLA radio images were also presented. There were two extraordinary findings. Firstly, we found using our detailed plasmoid analysis that it is likely the case that the long-term time-averaged jet power is among the largest known for a quasar, $\overline{Q} \approx 2.66 \pm 1.33 \times 10^{47} \rm{ergs/sec}$. This result is quantitatively similar to the cruder standard estimate that is computed using only the 151 MHz flux density, $\overline{Q} \approx 2.07 \pm 0.69 \times 10^{47} \rm{ergs/sec}$. It follows that 3C82 is the host of likely one of, if not, the most kinetically dominated known quasar jets. Namely, if the accretion flow bolometric luminosity is $L_{\rm{bol}}$ ($L_{\rm{bol}}\approx 3.2-5.8 \times 10^{46} \rm{ergs/sec}$ from our estimates of 3C 82) then the ratio of $\overline{Q}/L_{\rm{bol}}\approx 5.91 \pm 3.41$ (see the discussion in Section 7.2) is larger than the other known powerful jets (see Table 5). We also showed, in Section 7.2 and Table 3, that this result was likely true over a wide range of assumptions. It is true even if we do not minimize the residuals of the fit to the radio data (as we do in our preferred solution), but allow the fit to vary within the constraints of the error bars of the radio data. We also showed that $\overline{Q}$ is the same to within 10\% if deviations from the minimum energy assumption are allowed, or if the lobes advance at velocities $\leq 0.1$c. These are many compelling reasons to believe that our estimates are robust.

Secondly, the UV spectrum revealed strong evidence of a powerful high ionization outflow. We estimated $P_{w} \gtrsim 10^{45} \rm{ergs/sec}$ with two different wind models. Evidence of powerful high ionization outflows is commonly seen in both UV absorption and UV emission in high Eddington rate radio quiet quasars, but is very rare in radio loud quasars with an FRII morphology \citep{ric02,pun10,bec00,bec01}. Yet, 3C82 has a strong high ionization wind even though it has perhaps the most powerful jet. Why these two almost mutually exclusive properties exist in such a powerful jet system is a mystery and a valuable clue to the physical nature of the jet/wind launching mechanism. We have noticed that other 3C CSS quasars such as 3C 286 may also have this unusual property, for a radio loud quasar, of excess blue-shifted CIV emission. It is an interesting possibility that the immediate environment of the accretion flow is different in this subclass of powerful radio sources. We have obtained deep wide band optical spectra of 3C CSS quasars to augment the HST archives. We plan to perform a complete spectral analysis and report our findings in a future work. It is tempting to speculate an evolutionary scenario in which luminous 3C CSS quasars are the precursors to the large FRII radio sources \citep{ode98}. In that case, a prolonged period of high accretion rate, accompanied by powerful baryonic winds, may be a common (though relatively short-lived) phase that might be instrumental in the establishment of the long-term jet behavior.
\section{Acknowledgments}
This paper benefitted greatly from the report of a very knowledgable referee. We thank the staff of the Hobby Eberly Telescope for support of our LRS2 observations as well as the LRS2 commissioning team. The LRS2 spectrograph was funded by McDonald Observatory and Department of Astronomy, University of Texas at Austin, and the Pennsylvania State University. The supercompter Maverick hosted by the Texas Advanced Computing Center was used in the storage and reduction of the LRS2 data. This research was partially funded through McDonald Observatory. The Hobby-Eberly Telescope (HET) is a joint project of the University of Texas at Austin, the Pennsylvania State University, Ludwig-Maximilians-Universität München, and Georg-August-Universität Göttingen. The HET is named in honor of its principal benefactors, William P. Hobby and Robert E. Eberly. We would like to thank Anita Richards and Rob Beswick of the MERLIN/VLBI National Facility for supplying the 5 GHz data. This work was supported by the
National Radio Astronomy Observatory, a facility of the National Science Foundation operated under cooperative agreement by Associated Universities, Inc.


\begin{thebibliography}{}
\bibitem[Akujor and Garrington(1995)]{aku95} Akujor C. E., Garrington S. T., 1995, A\&AS, 112, 235
\bibitem[Alexander(2006)]{ale06} Alexnader, P. 2006 MNRAS, 368, 1404.
\bibitem[Alexander and Pooley(1996)]{ale96} Alexnader, P. and Pooley, G. 1996 in \emph{Cygnus A - Study of a Radio Galaxy}, eds. C. Carilli and D. Harris (Cambridge University Press, New York), 149
\bibitem[Baars et al.(1977)]{baa77} Baars, J., Genzel, R., Pailiny-Toth, I.I.K., Witzel, A., 1977, A\&A, 61, 99
\bibitem[Barthel(1989)]{bar89} Barthel, P. D. 1989, ApJ, 336, 606
\bibitem[Barthel et al.(1990)]{bar90} Barthel, P., Tytler, D., Thomson, B. 1990, A and A Supp. 82 339
\bibitem[Barthel and Arnaud(1996)]{bar96} Barthel, P. D.. Arnaud, K. 1996, MNRAS, 283, L45.
\bibitem[Bauman et al(2013)]{bau13}Baumann, G., Galsgaard, K. \& Norlund, A., 2013 Solar Physics 284 467
\bibitem[Becker et al.\ (2000)]{bec00} Becker., R. et al.\ 2000, ApJ, 538, 72
\bibitem[Becker et al.\ (2001)]{bec01} Becker., R. et al.\ 2000, ApJS, 135, 227
\bibitem[Bicknell et al.\ (1990)]{bic90} Bicknell, G., DeRuiter. H., Fanti, R., Morganti, R., Parma, P.\ 1990, ApJ, 354, 98
\bibitem[Blandford and K{\"o}ingl(1979)]{bla79}Blandford, R. and K{\"o}nigl, A. 1979, ApJ 232 34
\bibitem[Blundell and Rawlings(2000)]{blu00} Blundell, K., Rawlings, S. 2000 AJ 119 1111
\bibitem[Brotherton et al(1994)]{bro94}Brotherton, M., Wills B., Steidel, C., Sargent, W. 1994, ApJ 430
131
\bibitem[Brotherton(1996)]{bro96}Brotherton, M. 1996, ApJS 102 1
\bibitem[Cardelli et al.(1989)]{car89} Cardelli, J., Clayton, G., Mathis, J. 1989 ApJ 345
245
\bibitem[Chambers et al.(2019)]{cha19} Chambers, K., Magnier1, E. Metcalfe, N., Flewelling1, H. Huber, M. et al.\ 2019, arXiv:1612.05560
\bibitem[Chonis et al.(2016)]{cho16} Chonis, T.~S., Hill, G.~J., Lee, H., et al.\ 2016, \procspie, 9908, 99084C
\bibitem[Cohen et al.(2007)]{coh07} Cohen, A. S.; Lane, W. M.; Cotton, W. D.; Kassim, N. E.; Lazio, T. J. W. et al. 2007, AJ, 134, 1235
\bibitem[Condon et al.(1998)]{con98} Condon J. J., Cotton W. D., Greisen E. W., Yin Q. F., Perley R. A., Taylor G. B., Broderick J. J., 1998, AJ, 115, 1693
\bibitem[Croston et al.(2005)]{cro05} Croston J. H., Hardcastle M. J., Harris D. E., Belsole E., Birkinshaw M., Worrall D. M., 2005, ApJ, 626, 733
\bibitem[Croston et al.(2018)]{cro18} Croston, J. H.; Ineson, J.; Hardcastle, M. J, 2018, MNRAS, 476, 161
\bibitem[Davis and Laor(2011)]{dav11}Davis, S., Laor, A. 2011, ApJ 728 98
\bibitem[Evans and Koratkar(2004)]{eva04} Evans, I. and Koratkar 2004, ApJS 150 73
\bibitem[Fanaroff and Riley (1974)]{fr74}Fanaroff, B.~L.; Riley, J. ~M., 1974, MNRAS, 167, 31P
\bibitem[Fender et al.(1999)]{fen99}Fender, R. et al., 1999, MNRAS 304 865
\bibitem[Fernini(2007)]{fer07}Fernini, I., 2007, AJ 134 158
\bibitem[Fernini(2014)]{fer14}Fernini, I., 2014, ApJS 212 19
\bibitem[Fine et al.(2010)]{fin10}Fine, S., Croom, S., Bland-Hawthorne, J., Pimbblet.,K., Ross, N. 2010, MNRAS 409 591
\bibitem[Ghisellini et al.(2010)]{ghi10}Ghisellini, G, Tavecchio, F. and Foschini, L. et al. 2010 MNRAS
402 497
\bibitem[Ginzburg and Syrovatskii(1965)]{gin65} Ginzburg, V. and Syrovatskii, S. 1965,
  Annu. Rev. Astron. Astrophys. 3 297
\bibitem[Ginzburg and Syrovatskii(1969)]{gin69} Ginzburg, V. and Syrovatskii, S. 1969,
  Annu. Rev. Astron. Astrophys. 7 375
\bibitem[Gower et al.(1967)]{gow67}Gower, J., Scott, P., Wills, P. 1967 MNRAS 71 49
\bibitem[Hales et al.\ (1993)]{hal93}Hales, S., Baldwin,J., Warner, P. 1993 MNRAS 263 25
\bibitem[Hardcastle and Worrall(2000)]{har00} Hardcastle, M., Worrall, D. 2000, MNRAS,
314, 359
\bibitem[Hardcastle et al.(2004)]{har04} Hardcastle, M. J.; Harris, D. E.; Worrall, D. M.; Birkinshaw, M. 2004, ApJ, 612, 729
\bibitem[Hardcastle et al.(2009)]{har09} Hardcastle, M., Evans, D. and Croston, J. 2009, MNRAS, 396,
1929
\bibitem[Hill et al.\ (2018)]{hill18} Hill, G.~J., Drory, N., Good, J.~M., et al.\ 2018b, \procspie, 10700, 107000P
\bibitem[Homan et al.\ (2002)]{hom02}Homan, D. C., Ojha, R., Wardle, J. F. C., Roberts, D. H., Aller, M. F., Aller,
H. D., \& Hughes, P. A. 2002, ApJ, 568, 99
\bibitem[Hurley-Walker(2017)]{hur17} Hurley-Walker, N.\ 2017 arXiv:1703.06635
\bibitem[Intema et al.(2017)]{int17} Intema, H. T., Jagannathan, P., Mooley, K. P., \& Frail, D. A. 2017, A\&A, 598, A78
\bibitem[Kaiser and Alexander(1997)]{kai97}Kaiser C. R., Alexander P., 1997, MNRAS, 286, 215
\bibitem[Kanekar et al.(2009)]{kan09} Kanekar N., Lane W. M., Momjian E., Briggs F. H., Chengalur J. N., 2009, MNRAS  394 L61
\bibitem[Kanekar et al.(2013)]{kan13} Kanekar N., Ellison, S.., Momjian E., York B., Pettini N., 2013, MNRAS 428 532
\bibitem[Kataoka and Stawartz(2005)]{kat05}Kataoka J., Stawarz Ł., 2005, ApJ, 622, 797
\bibitem[Kharb et al.(2008)]{kha08} Kharb, P., O'Dea, C. Baum, S. et al. 2008 ApJS 174, 74
\bibitem[Krongold et al.(2003)]{kro03}Krongold, Y., Nicastro, F., Brickhouse, N. S., Elvis, M, Liedahl, D., Mathur, S. 2003 ApJ, 597, 832
\bibitem[K{\"u}hr et al.(1981)]{kuh81}K{\"u}hr, H., Witzel, A., Pauliny-Toth, I.I.K., Nauber, U.1981 A\&AS, 45, 367
\bibitem[Laing and Peacock(1980)]{lai80}Laing R. A., and Peacock J.A., 1980, MNRAS, 190, 93
\bibitem[Laing et al.(1983)]{lai83}Laing R. A., Riley J. M., Longair M. S., 1983, MNRAS, 204, 15
\bibitem[Lane et al.(2012)]{lan12} Lane, W. M., Cotton, W. D., Helmboldt , J. and Kassim, N. E. 2012 "VLSS Redux: Software Improvements applied to the Very Large Array Low-frequency Sky Survey", Radio Science v. 47, RS0K04
\bibitem[Laor et al.(1997)]{lao97} Laor, A. Fiore, F., Elvis, M., Wilkes, B., McDowell, J. (1997) ApJ 477 93
\bibitem[Laor and Davis(2014)]{lao14}Laor, A., Davis, S. 2014 ApJ 428 3024
\bibitem[Lightman et al.(1975)]{lig75}Lightman, A., Press, W., Price, R. and Teukolsky, S. 1975, \emph{Problem Book in Relativity and Gravitation} (Princeton University Press, Princeton)
\bibitem[Lind and Blandford(1985)]{lin85}Lind, K., Blandford, R.
1985, ApJ 295 358
\bibitem[Liu et al.(1992)]{liu92}Liu R., Pooley G. G., Riley J. M., 1992, MNRAS, 257, 545
\bibitem[Lira et al.(2018)]{lir18}Lira, P., Kaspi, S., Netzer, H. 2018, ApJ 865 56
\bibitem[Ludke et al.(1998)]{lud98}Ludke, E., Garrington S., Spencer R., et al., 1998, MNRAS, 299, 467
\bibitem[Malakit(2009)]{mal09}Malakit, K., Cassak, P., Shav, M., Drake, F. 2009, Geosphysical Research Letter 36 L07107
\bibitem[Malkan(1983)]{mal83} Malkan, M. 1983, ApJ 268, 582
\bibitem[Marziani et al.(1996)]{mar96}Marziani, P., Sulentic, J., Dultzin-Hacyan, D., Calvani, M., Moles, M. 1996, ApJS 104 37
\bibitem[Marziani et al.(2010)]{mar10}Marziani, P., Sulentic, J., Negrete, A., Dultzin-Hacyan, D., Zamfir, S. al. 2010, MNRAS 409 1033
\bibitem[Marziani et al.(2017)]{mar17}Marziani, P., Negrete, A., Dultzin, D. et al. 2017, Frontiers in Astronomy and Space Sciences, Volume 4, id.16
\bibitem[Mathews and Brighenti(2003)]{mat03}Mathews, W. and Brighenti, F. 2003, ARA \& A 41 191
\bibitem[McNamara et al.(2011)]{mcn11}McNamara, B., Rohanizadegan, M, and Nulsen, P. 2011 ApJ 727
39
\bibitem[Moffet(1975)]{mof75}Moffet, A. 1975 in \emph{Stars and Stellar Systems, IX: Galaxies and the Universe},
eds. A. Sandage, M. Sandage \& J. Kristan (Chicago University Press,
Chicago), 211.
\bibitem[Murray et al.(1995)]{mur95} Murray, N., Chiang J. Grossman, S, and Voit, G. 1995, ApJ 451, 498
\bibitem[Nandra et al.(1997)]{nan97} Nandra, K., George, I. M., Mushotzky, R. F., Turner, T. J. and  Yaqoob, T.
1997 ApJ 476 30
\bibitem[Netzer and Marziani\ (2010)]{net10} Netzer, H., and Marziani, P. 2010, ApJ 724, 318
\bibitem[Novikov and Thorne(1973)]{nov73} Novikov, I. and Thorne, K. 1973, in \emph{Black Holes: Les Astres Occlus}, eds. C. de Witt and B. de Witt (Gordon and
Breach, New York), 344
\bibitem[O'Dea(1998)]{ode98} O'Dea, C. 1998, PASP, 110, 493
\bibitem[Orienti and Dallacasa (2008)]{ori08} Orienti, M. and Dallacasa, D. 2008, A \& A 487 885
\bibitem[Owen et al.(2000)]{owe00}Owen F. N., Eilek J. A., Kassim N. E., 2000, ApJ, 543 611
\bibitem[Parker\ (1958)]{par58}Parker E. N., 1958, ApJ, 128, 664
\bibitem[Pearson et al.\ (1985)]{pea85}Pearson T. J., Readhead A. C. S., Perley R. A., 1985, AJ, 90, 738
\bibitem[Perley and Butler(2013)]{per13}Perley, R. and Butler, B. 2013, ApJS 204 19
\bibitem[Punsly(2006)]{pun06}Punsly, B.\ 2006, ApJ, 647, 886
\bibitem[Punsly(2008)]{pun08} Punsly, B. 2008, \emph{Black Hole Gravitohydromagnetics}, second edition (Springer-Verlag, New York)
\bibitem[Punsly(2010)]{pun10}Punsly, B. 2010, ApJ 713 232
\bibitem[Punsly(2012)]{pun12} Punsly, B. 2012, ApJ 746 91
\bibitem[Punsly(2015)]{pun15}Punsly, B. 2015 ApJ 806 47
\bibitem[Punsly et al.(2016)]{pun16}Punsly, B., Marziani, P., Zhang, S., Muzahid, S., O'Dea, C. 2016 ApJ 830 104
\bibitem[Punsly et al.(2018)]{pun18}Punsly, B., Tramacere, P., Kharb, P., Marziani, P. 2018 ApJ 869 164
\bibitem[Punsly(2019)]{pun19}Punsly, B. 2019 ApJL 871 34
\bibitem[Rawlings et al.(1989)]{raw89}Rawlings, S., Eales, S. Riley, J. and Saunders, R. 1989 MNRAS 240 723
\bibitem[Rees(1966)]{ree66} Rees, M. J. 1966, Nature 211: 468-70
\bibitem[Readhead et al.(1996)]{rea96}Readhead A. C. S., Taylor G. B., Xu W., Pearson T. J., Wilkinson P. N.,Polatidis A. G., 1996, ApJ, 460, 612
\bibitem[Reid et al.(1995)]{rei95}Reid, A., Shone, D., Akujor, C., et al. 1995, A\%AS.
110 213
\bibitem[Reynolds et al.(2009)]{rey09}Reynolds, C., Punsly, B. Kharb, P., O'Dea, C. and Wrobel, J. 2009, ApJ, 706, 851
\bibitem[Reynolds et al.(2020)]{rey20} Reynolds, C., Punsly, B., Miniutti, G., O'Dea, C., and Hurley-Walker, N. 2020 to appear in ApJ arXiv:2001.10697
\bibitem[Richards et al(2002)]{ric02}Richards, G. et al 2002, AJ 124 1
\bibitem[Riley(1989)]{ril89}Riley, J. 1989, MNRAS 238,1055
\bibitem[Scheuer(1995)]{sch95}Scheuer P. A. G., 1995, MNRAS, 277, 331
\bibitem[Semenov et al.(2004)]{sem04}Semenov, V. Dyadechkin, S., Punsly, B. 2004, Science 305 978
\bibitem[Sulentic et al.(2000)]{sul00}Sulentic, J., Marziani, P., and Dultzin-Hacyan, D. 2000 ARA\& A 38, 521
\bibitem[Sulentic et al.(2007)]{sul07}Sulentic, J., Bachev, R., Marziani,P., Negrete, C. A., Dultzin, D. 20007
ApJ 666 757
\bibitem[Sulentic et al.(2015)]{sul15}Sulentic, J., Martinez-Caraballo, M., Marziani, P. et al. 2015 MNRAS 450 1916
\bibitem[Sulentic et al.(2017)]{sul17}Sulentic, J.,del Olmo., A., Marziani, P. et al. 2017 A \& A 608 122
\bibitem[Taylor and Perley\ (1992)]{tay92}Taylor, G. and Perley, R.\ 1992, A\& A, 262 417.
\bibitem[Telfer et al.(2002)]{tel02} Telfer, R., Zheng, W., Kriss, G., Davidsen, A. 2002 ApJ 565 773
\bibitem[Threlfall et al.(2012)]{thr12} Threlfall, J. et al 2012, A \& A 544 24
\bibitem[Tucker(1975)]{tuc75}Tucker, W. 1975, \emph{Radiation Processes in
    Astrophysics} (MIT Press, Cambridge).
\bibitem[van Breugel et al.(1992)]{van92} van Breugel W. J. M., Fanti C., Fanti R., Stanghellini C., Schilizzi R. T.,
Spencer R. E., 1992, A\&A, 256, 56
\bibitem[van der Laan(1966)]{van66} van der Laan, H. 1966, Nature 211 1131
\bibitem[Walker et al.(1987)]{wal87}Walker, R., Benson, J., Unwin, S. 1987 ApJ 316 546
\bibitem[Weymann et al.(1991)]{wey91} Weymann, R.J., Morris, S.L., Foltz,
      C.B., Hewett, P.C. 1991, ApJ, 373, 23
\bibitem[Weymann(1997)]{wey97}Weymann, R. 1997 in ASP Conf. Ser. 128, Mass
      Ejection from Active Nuclei ed, N.Arav, I. Shlosman and R.J. Weymann (San
      Francisco: ASP), 3
\bibitem[Willott et al.(1999)]{wil99}Willott, C., Rawlings, S., Blundell, K., Lacy, M. 1999, MNRAS 309 1017
\bibitem[Wright(2006)]{wri06} Wright, E. L. 2006, PASP, 118, 1711
\bibitem[Yamada(2007)]{yam07} Yamada, M.2007, Physics of Plasmas 14 058102
\bibitem[York et al.(2007)]{yor07}York B. A., Kanekar N., Ellison S. L., Pettini M., 2007, MNRAS 382 L53
\bibitem[Zheng et al.(1997)]{zhe97} Zheng, W. et al. 1997 ApJ 475 469
\end{thebibliography}
\end{document}